\documentclass[aps,pre,reprint,groupedaddress, superscriptaddress]{revtex4-1}

\usepackage{graphicx}
\usepackage{amssymb}
\usepackage{amsmath}
\usepackage{bm}
\usepackage{hyperref}
\usepackage{lipsum}
\usepackage{bbold}
\usepackage{comment}

\begin{document}

\title{Nearly flat bands in twisted triple bilayer graphene}

\author{Jiseon Shin}
\affiliation{Department of Physics, University of Seoul, Seoul 02504, Korea}
\author{Bheema Lingam Chittari}
\affiliation{Department of Physical Sciences, Indian Institute of Science Education and Research Kolkata, Mohanpur 741246, West Bengal, India}
\author{Yunsu Jang}
\affiliation{Department of Physics and Astronomy, Seoul National University, Seoul 08826, Republic of Korea}
\author{Hongki Min}
\affiliation{Department of Physics and Astronomy, Seoul National University, Seoul 08826, Republic of Korea}
\author{Jeil Jung}
\email[jeiljung@uos.ac.kr]{}
\affiliation{Department of Physics, University of Seoul, Seoul 02504, Korea}
\affiliation{Department of Smart Cities, University of Seoul, Seoul 02504, Korea}

%\date{}

\begin{abstract}
We investigate the electronic structure of alternating-twist triple Bernal-stacked bilayer graphene (t3BG) as a function of interlayer coupling $\omega$, twist angle $\theta$, interlayer potential difference $\Delta$, and top-bottom bilayers sliding vector $\boldsymbol{\tau}$ for three possible configurations AB/AB/AB, AB/BA/AB, and AB/AB/BA. The parabolic low-energy band dispersions in a Bernal-stacked bilayer and gap-opening through a finite interlayer potential difference $\Delta$ allows the flattening of bands in t3BG down to $\sim 20$~meV for twist angles $\theta \lesssim 2^{\circ}$ regardless of the stacking types. The easier isolation of the flat bands and associated reduction of Coulomb screening thanks to the intrinsic gaps of bilayer graphene for finite $\Delta$ facilitate the formation of correlation-driven gaps when it is compared to the metallic phases of twisted trilayer graphene under electric fields. We obtain the stacking dependent Coulomb energy versus bandwidth $U/W \gtrsim 1$ ratios in the $\theta$ and $\Delta$ parameter space. We also present the expected $K$-valley Chern numbers for the lowest-energy nearly flat bands.
\end{abstract}

%\pacs{68.37.Ef, 82.20.-w, 68.43.-h}
%\keywords{Suggested keywords}

\maketitle

%---------------------------------------------------
\section{Introduction}
\begin{figure*}
\begin{center}
\includegraphics[width=0.95\textwidth]{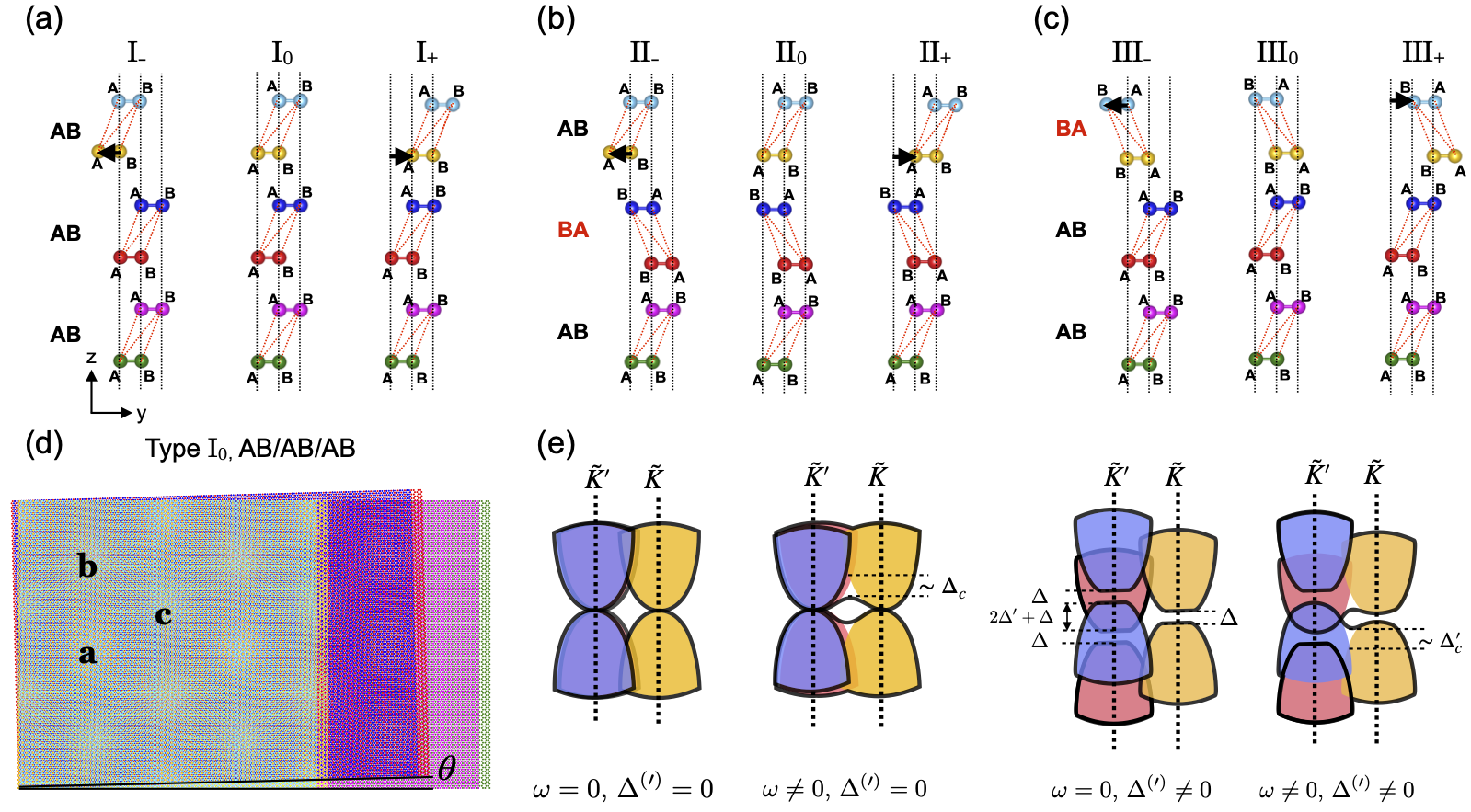}\\
\end{center}
\caption{
(Color online) Schematic diagrams for the starting stacking configurations before a twist of (a) type I$_{(-, 0, +)}$, (b) type II$_{(-, 0, +)}$, and (c) type III$_{(-, 0, +)}$. (d) The moire patterns at $\theta = 1.5^\circ$ in real space for type I$_0$ and the location of the corresponding three local commensurate stackings {\bf a}, {\bf b}, and {\bf c} are indicated. (e) Schematic diagrams for four different cases where the couplings between bilayers $\omega$ are considered as a perturbation. The three parabolic bands have the intrinsic bilayer gap ($\Delta$) and the two parabolic bands at $\tilde{K}'$ have the band offset (2$\Delta^{\prime}$+$\Delta$) due to an external electric field in the $z$-direction.  The three parabolic dispersive bands for the second ($\omega \neq 0$, $\Delta^{(\prime)} = 0$) and the fourth ($\omega \neq 0$, $\Delta^{(\prime)} \neq 0$) cases are plotted with dotted lines as a reference in Fig.~\ref{fig2}. When the tunneling is considered perturbatively, the interband hybridizations have $\Delta_c \approx 7.9$ meV and $\Delta_c' \approx 5.0$ meV at $\theta = 1.5^\circ$ for $\omega = 0.01$ eV and $\Delta = \Delta' = 0.01$ eV.
}
\label{fig1}
\end{figure*}

Intense research activity on the physics of twisted structures has unfolded since experimental observations of correlated phases and flat band superconductivity near its magic-angle in twisted bilayer graphene (t2G) \cite{Kim2017, Cao2018a, Cao2018b, Yankowitz2018a, Cao2019} and twisted trilayer graphene (t3G) ~\cite{Park2021, hao2020electric}. Many conceivable combinations of twisted graphene systems beyond t2G have been studied, 
including the twisted double bilayer graphene (t2BG)~\cite{Lee2019, Chebrolu2019, Choi2019,  Koshino2019, Burg2019, Cao2019a, Shen2019, Liu2019}
that shows electric-field-tunable lowest-energy bands up to two times narrower 
than that of t2G for similar system parameters \cite{Lee2019, Chebrolu2019, Choi2019, Koshino2019, Burg2019, Cao2019a, Shen2019, Liu2019}, and
the ABC-stacked trilayer graphene on hexagonal boron nitride~\cite{Chittari2019} that shows electric field tunable bandwidths and band isolation by opening a gap at the charge neutrality point. 
%
%
%The gaps in the constituent moire materials either intrinsic or opened through an electric field like in Bernal bilayer graphene or ABC-stacked trilayer graphene 
%facilitates the flattening and isolation of the low energy bands and relaxes the need of achieving very precise magic angles~\cite{srivani}. 
%
%
Other trilayer systems that have been studied recently include the twisted mono-bilayers~\cite{SuarezMorell2013,Li2019,szendr2020ultraflat,Carr2019a,Ma2019,tmbg1,tmbg2,tmbg3,Park2020, lei2020mirror, wu2020lattice}, 
the consecutive-twist \cite{Mora2019} and alternating twisted trilayer graphene (t3G)~\cite{Park2021, Zuo2018,Khalaf2019, Carr2019a, Li2019, tb_tTG,lei2020mirror, hao2020electric, Dumitru2021, Shin2021, Mora2019}.
%
%Concerning t3G possessing two moire interfaces, there have been studies of consecutive-twist \cite{Mora2019} and alternating-twist \cite{Park2021, Zuo2018,Khalaf2019, Carr2019a, Li2019, tb_tTG,lei2020mirror, hao2020electric, Dumitru2021, Shin2021}. It has been stated that the consecutive-twist trilayer graphene has a perfect metallic phase protected by particle-hole symmetry which does not show the nearly flat bands. The alternating-twist case, on the other hand, has nearly flat bands at lowest-energy regime at the magic-angle which is larger than that of t2G by $\sqrt{2}$. Besides, there are 2D Dirac cone-like linear dispersing bands that do not hybridize with the lowest-energy flat bands at zero electric field. Recently, alternating t3G was reported to violate the positive correlation between the density of states and the critical temperature at large electric field, to manifest ultra-strong coupling superconductivity, and to reach the BCS-BEC crossover \cite{Park2021, hao2020electric}.

%
%In this work we % accordance with these expansible developments such as doubling layers and adding one more moire interface, in this paper, 
This work expands the ongoing effort in search of graphene-based flat band systems 
by proposing that the electronic structure of twisted triple bilayer graphene (t3BG) formed by three Bernal-stacked bilayer graphene with alternating twists can offer advantages thanks to the electric field tunable intrinsic gaps of each bilayer graphene~\cite{min2007} and the possibility of achieving narrow bandwidths at larger twist angles due to the effective increase of the interlayer tunneling strength expected in alternating twist triple layer systems~\cite{tb_tTG}.
We perform numerical calculations for sufficiently small twist angles $\theta \leq 2.5 ^\circ$ in search of nearly flat low-energy bands by using
continuum moire bands models used in earlier works~\cite{Bistritzer2010b, Jung2014}.
A total of 9 different starting stacking combinations are considered, 
to account for the three different possible stacking orientations and three sliding 
positions of the top bilayer with respect to the bottom bilayer,
%Given that the lattice is relaxed, giving rise to the imbalance between the intra- and inter-sublattice as observed in several experiments \cite{Carr2019, Cao2016, Cao2018a}, 
and we consider the remote hopping terms in the bilayer graphene~\cite{Jung2014a} for a more accurate description of the electronic structure. 
%to describe the system in touch with reality, and we call it remote hopping model for short. Note that the nearly flat bands of our interest are distinct from completely flat bands in the momentum space, which are describable using compact localized states \cite{Derzhko2015, Leykam2018, Leykam2018a}. 
%
%
Our phase diagram calculations in the parameter space of twist angle $\theta$ and perpendicular electric fields induced by interlayer potential differences $\Delta$ indicate that narrow bandwidths below $\sim 20$~meV can be achieved for twist angles below $\theta \lesssim 2^{\circ}$ regardless of stacking types. A perpendicular electric field can be used as a control knob of the bandwidths and band isolation that can facilitate the onset of correlation gaps and valley Chern bands.

Our manuscript is organized as follows. In Sec~\ref{Model}, we give a detailed description of the stacking configurations at each moire interface and the associated model Hamiltonian. We devote Sec.~\ref{R1} to describe how the electronic properties of t3BG change as a function of the twist angles, interlayer couplings, starting stackings, sliding of the top bilayer, and electric fields. In Sec.~\ref{R2}, we explore a wide parameter space of the interlayer potential differences and twist angles in search of large Coulomb energy versus bandwidth ($U/W >1$) regime.
In Sec.~\ref{R3} we describe the valley Chern number phase diagrams that characterizes the topological properties of the lowest-energy electron and hole bands in t3BG. We close the paper in Sec.~\ref{Summary} with the conclusions. 

%---------------------------------------------------
\section{Model}\label{Model}

\begin{figure*}
\begin{center}
\includegraphics[width=1.0\textwidth]{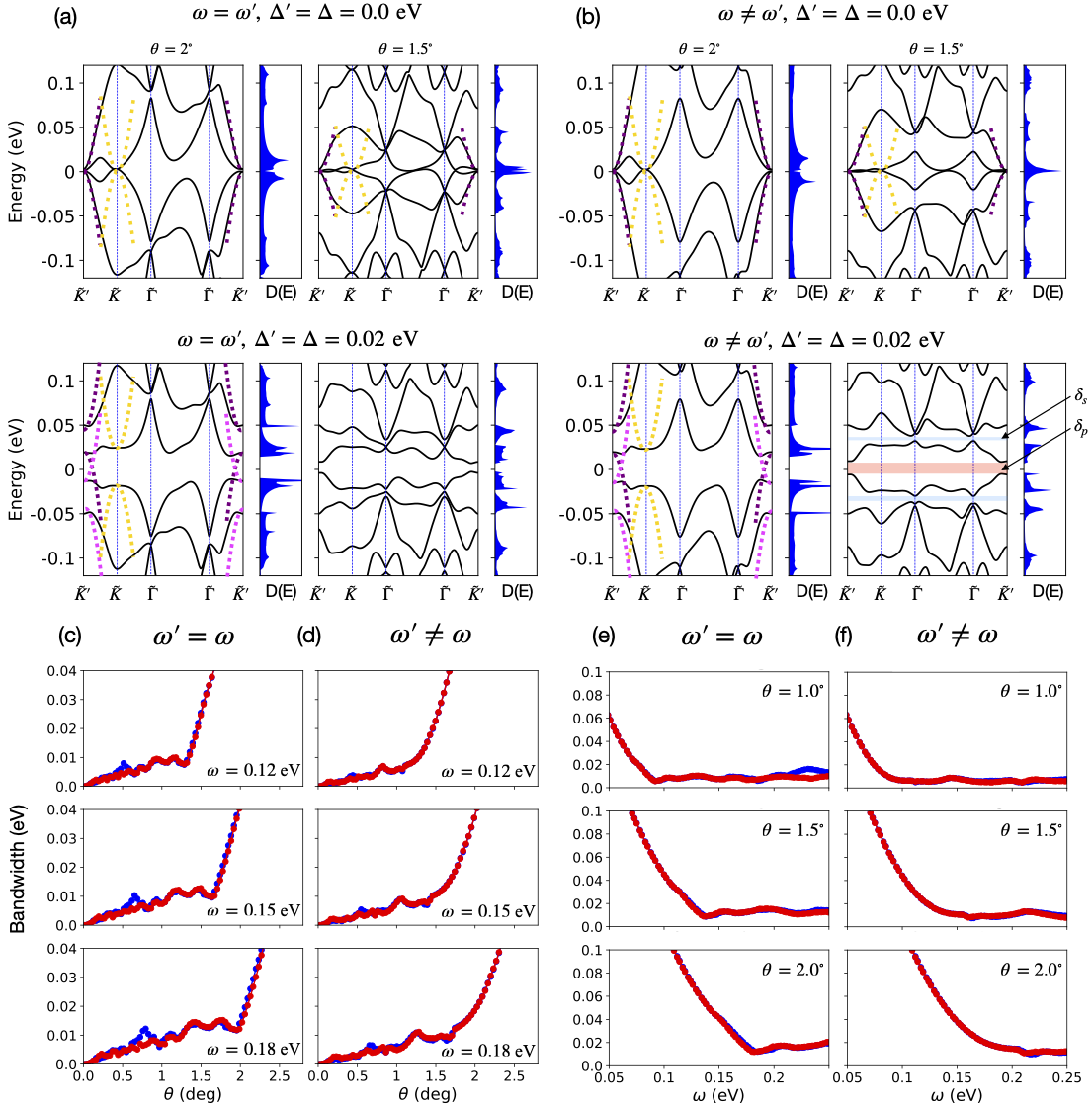}
\end{center}
\caption{
(Color online) Energy dispersions and density of states (DOS) of t3BG for type I$_0$ at $\theta = 2^\circ$ and $\theta = 1.5^\circ$ for (a) rigid ($\omega = \omega'$) and (b) relaxed ($\omega \neq \omega'$) lattice where $\Delta~(\Delta')$ represents the displacement field within (between) bilayers. The parabolic bands due to decoupled bilayers in Fig.~\ref{fig1} (e) which are indicated by the purple and yellow dotted lines in this figure are superposed on the energy dispersions. For zero interlayer displacement field, the effective mass of one of the two parabolic dispersions at $\tilde{K}^\prime$ remains constant regardless of the twist angle, while the other one at $\tilde{K}^\prime$ and the one at $\tilde{K}$ hybridize with each other and generate the nearly flat bands at the magic-angle. For finite interlayer displacement field, the three parabolic dispersions are depicted with pink, purple, and yellow dotted lines and note that the effective mass of the two parabolic dispersions at $\tilde{K}^\prime$ varies upon turning on the interlayer displacement field. Bandwidths of the lowest-energy valence (red) and conduction (blue) bands in type I$_0$ as a function of the twist angle $\theta$ for three different interlayer tunnelings $\omega = 0.12, 0.15, 0.18$~eV when (c) $\omega = \omega'$ and (d) $\omega \neq \omega'$, and as a function of $\omega$ for three different twist angles $\theta = 1.0^\circ, 1.5^\circ, 2^\circ$ when (e) $\omega = \omega'$, (f) $\omega \neq \omega'$ such that $\omega' = C_2 \omega^2 + C_1 \omega + C_0$ where $C_2 = -0.5506$ eV$^{-1}$, $C_1 = 1.036$, and $C_0 = -0.02245$ eV [see Sec.  \ref{Model}]. 
}\label{fig2}
\end{figure*}

\begin{figure*}
\begin{center}
\includegraphics[width=1\textwidth]{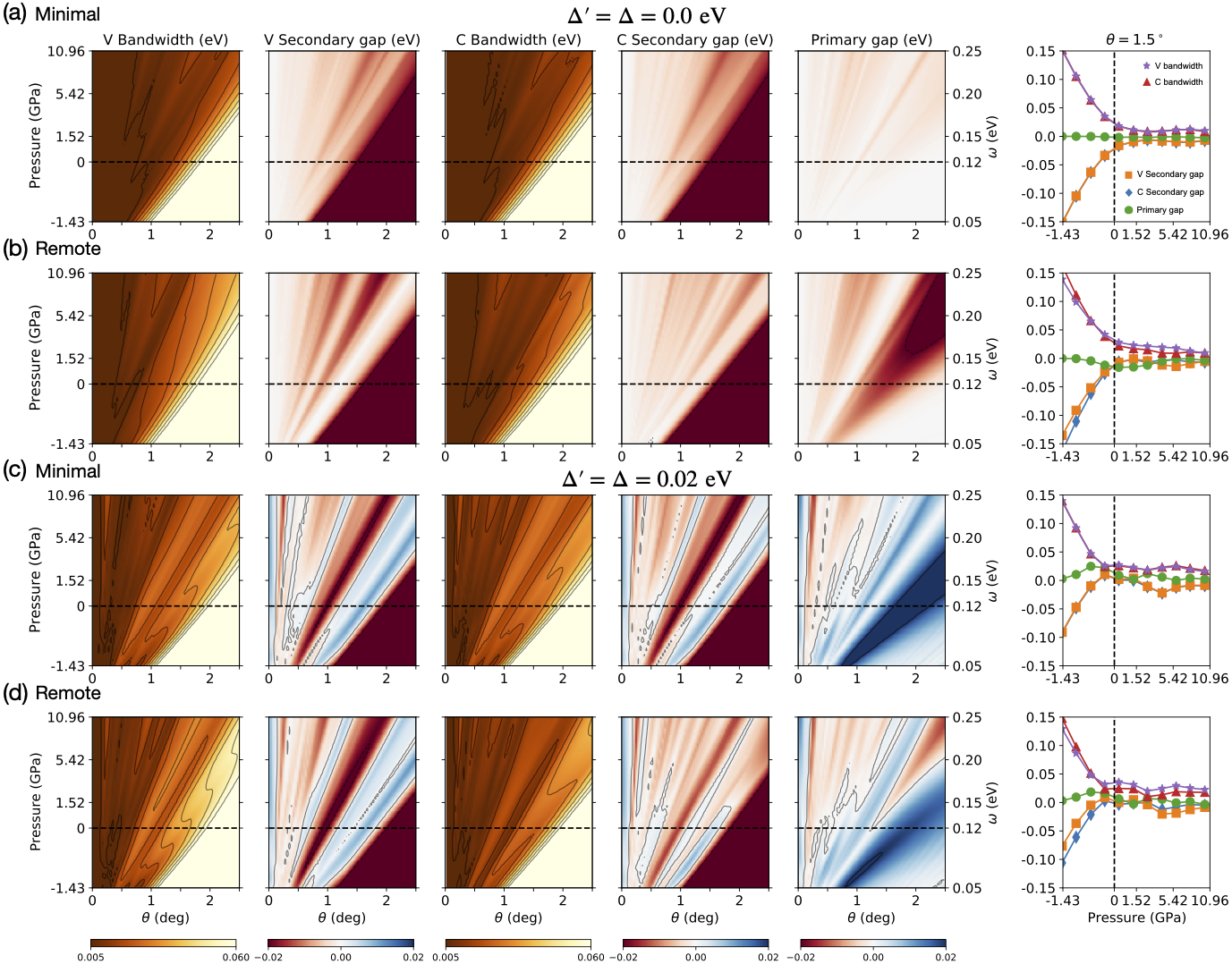}
\end{center}
\caption{
(Color online) Bandwidth, secondary bandgap, and primary bandgap of the lowest-energy valence (V) and conduction (C) bands of type I$_0$ for $\Delta' = \Delta = 0.0$ eV for the (a) minimal and (b) remote hopping models, and $\Delta' = \Delta = 0.02$ eV for the (c) minimal and (d) remote hopping model as a function of the pressure and the twist angle where $\Delta~(\Delta')$ represents the gate potential difference within (between) bilayers. The solid lines in the first and the third columns represent contours of the bandwidth of the lowest-energy valence and the conduction bands. The solid lines in the second, fourth and fifth columns represent the contour at the zero gaps. The cross-section at $\theta = 1.5^\circ$ is presented on the rightmost column. The dashed lines indicate zero pressure. 
}\label{fig3}
\end{figure*}

We consider three vertically stacked graphene bilayers with alternating twists such that the top and bottom bilayers are rotated by $-\theta/2$ (clockwise), while the middle bilayer is rotated by $+\theta/2$ (counterclockwise) as shown in Fig.~\ref{fig1}. We note that this twist configuration leading to commensurate moire patterns is favored in experimental devices~\cite{Park2021, hao2020electric, Park2021b}. We distinguish three starting stacking configurations AB/AB/AB, AB/BA/AB, and AB/AB/BA and call them type I$_0$, II$_0$, and III$_0$ respectively. 
Subsequently, we take into account slidings of the top bilayer with respect to the bottom bilayer by $0$, $\pm a/\sqrt{3}$ along the $y$-axis parallel to the carbon dimers in the unit cell which we express compactly as $\bm{\tau}_{0, \pm}$. 
We note that $\pm 3\bm{\tau}_\pm$ translation brings the system back to the original configuration. 
Also, a sliding by $\bm{\tau}_+$ ($\bm{\tau}_-$) is equivalent to $-2\bm{\tau}_-$ ($+2\bm{\tau}_+$). Therefore, we consider 9 in total possible starting stacking geometries that we name as 
I$_{(0, +, -)}$, II$_{(0, +, -)}$, and III$_{(0, +, -)}$ as shown in Figs.~\ref{fig1}(a)-(c) where 
a translation of the top bilayer by $\bm{\tau}_0$ or $\bm{\tau}_\pm$ is denoted by $0$ or $\pm$ in the subscript of the stacking types, I, II or III. 
The moire patterns for type I$_0$ are illustrated in Fig.~\ref{fig1}(d) where we indicate the three representative local commensurate stacking configurations by the letters {\bf a}, {\bf b}, {\bf c}. 
All 9 starting stacking geometries in Figs.~\ref{fig1}(a)-(c) correspond to the local commensurate stacking at location {\bf a} in the moire patterns where we put the rotation axis. 
We will show in Sec.~\ref{R1} how the electronic band structures and local density of states are modified depending on the initial stacking geometries. 
\begin{figure*}
\begin{center}
\includegraphics[width=1.0\textwidth]{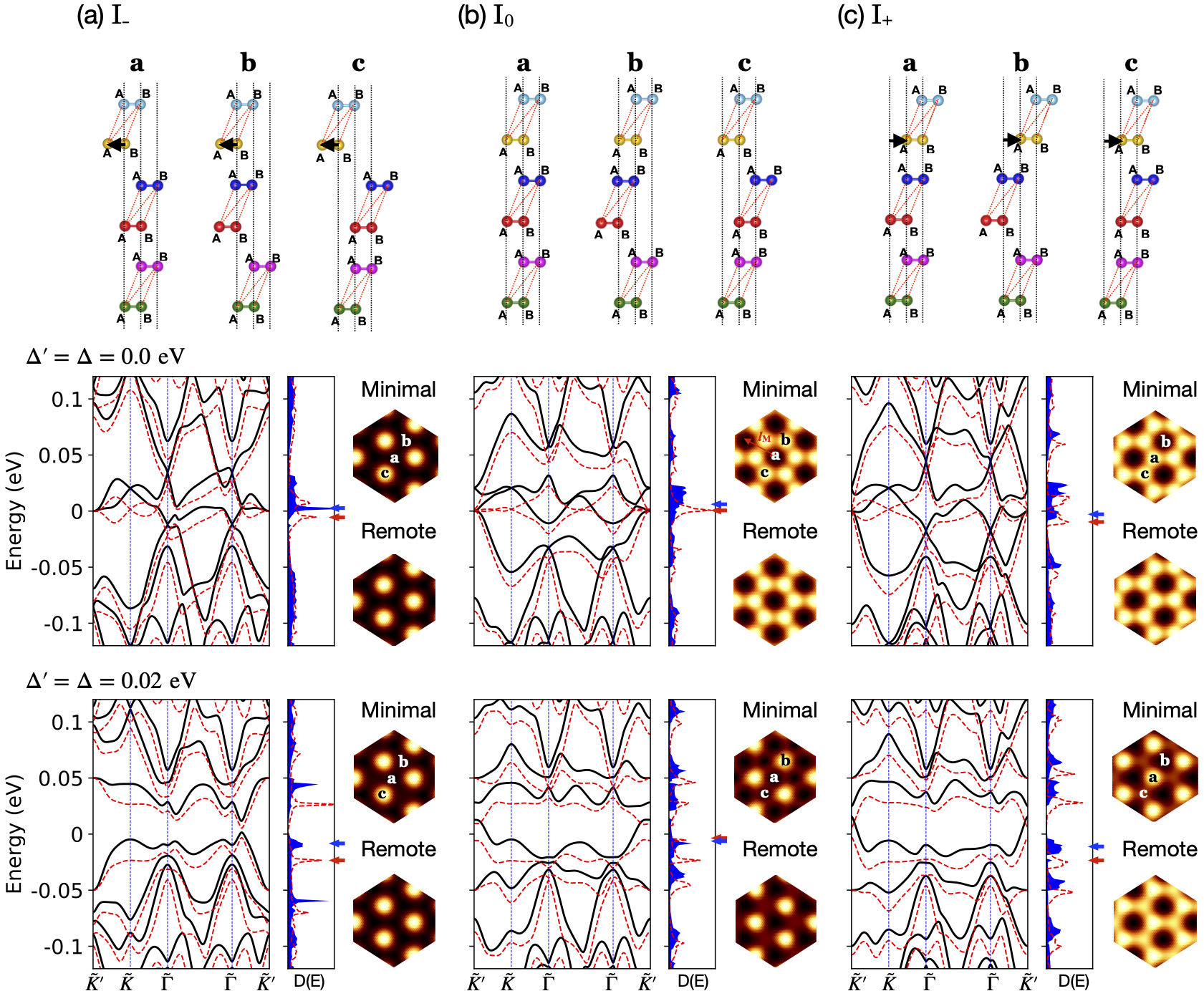}
\end{center}
\caption{
(Color online) Schematic diagrams for the atomic configurations of local commensurate stackings at {\bf a}, {\bf b}, and {\bf c} on the upper row, the band structures and the associated DOS at $\theta = 1.5^\circ$ for the minimal model (red dotted lines for bands and DOS) and for the remote model (black solid lines for bands, and blue filled area for DOS) with LDOS in the hexagonal frames at the van Hove singularities denoted by the red (minimal model) and the blue (remote model) arrows,
when $\Delta' = \Delta = 0.0$~eV 
in the middle row, and when $\Delta' = \Delta = 0.02$~eV in the lower row for 
(a) type I$_-$, (b) type I$_0$, (c) type I$_+$.
}\label{fig4}
\end{figure*}
The Hamiltonian of our model can be expressed concisely as
\begin{widetext}
\begin{equation}
H_\textrm{t3BG} (\theta) = \left(\begin{array}{cccccc}
h_{l,b}^- + \bar{V}_1 & t_s^- & 0 & 0 & 0 & 0 \\
t_s^{-\dagger} & h_{u,b}^- + \bar{V}_2 & T_1(\boldsymbol{r}) & 0 & 0 & 0\\
0 & T_1^{\dagger}(\boldsymbol{r}) & h_{l,m}^+ + \bar{V}_3 & t_s^+ & 0 & 0\\
0 & 0 & t_s^{+\dagger} & h_{u,m}^+ + \bar{V}_4 & T_2(\boldsymbol{r}) & 0\\
0 & 0 & 0 & T_2^{\dagger}(\boldsymbol{r}) & h_{l,t}^- + \bar{V}_5 & t_s^- \\
0 & 0 & 0 & 0 & t_s^{-\dagger} & h_{u,t}^- + \bar{V}_6 \\
\end{array}\right),
\end{equation}
\end{widetext}
%%%%%%%%%%% GJ %%%%%%%%%%%%%
%
where $h^{\pm}_{r, (b, m, t)} = h(\pm \theta/2) + (\mathbb{1} + r m_s \sigma_z) \hspace{0.05cm} \delta/2$ for the upper ($r = +1$) or the lower ($r = -1$) layer within an AB-stacked ($m_{s = {\rm AB}} = 1$) or 
BA-stacked ($m_{s = {\rm BA}} = -1$) bilayer located at the bottom (b), middle (m), or top (t). 
The first term represents a $2\times2$ Dirac Hamiltonian 
$h(\theta = 0) = {\upsilon}_F \boldsymbol{p} \cdot \boldsymbol{\sigma}$ rotated by $\pm \theta/2$ which is given by
\begin{equation}
h(\pm \theta/2)= D^\dagger (\pm \theta/2) ~ h(\theta = 0) ~ D (\pm \theta/2),
\end{equation}
where $D(\phi) = e^{-i \phi \sigma_z /2}$ is the general form of the spin $1/2$ rotation operator with the $z$-component Pauli matrix $\sigma_z$, and $\upsilon_F = \sqrt{3} \vert t_0 \vert a/ 2\hbar$, where we choose $t_0 = -3.1$~eV. 
The second term describes the intrinsic sublattice potential $\delta = 0.015$~eV due to the Bernal-stacked bilayer emerging at higher-energy dimers at the upper or lower layers~\cite{Jung2014a}.

\begin{figure*}
\begin{center}
\includegraphics[width=1.0\textwidth]{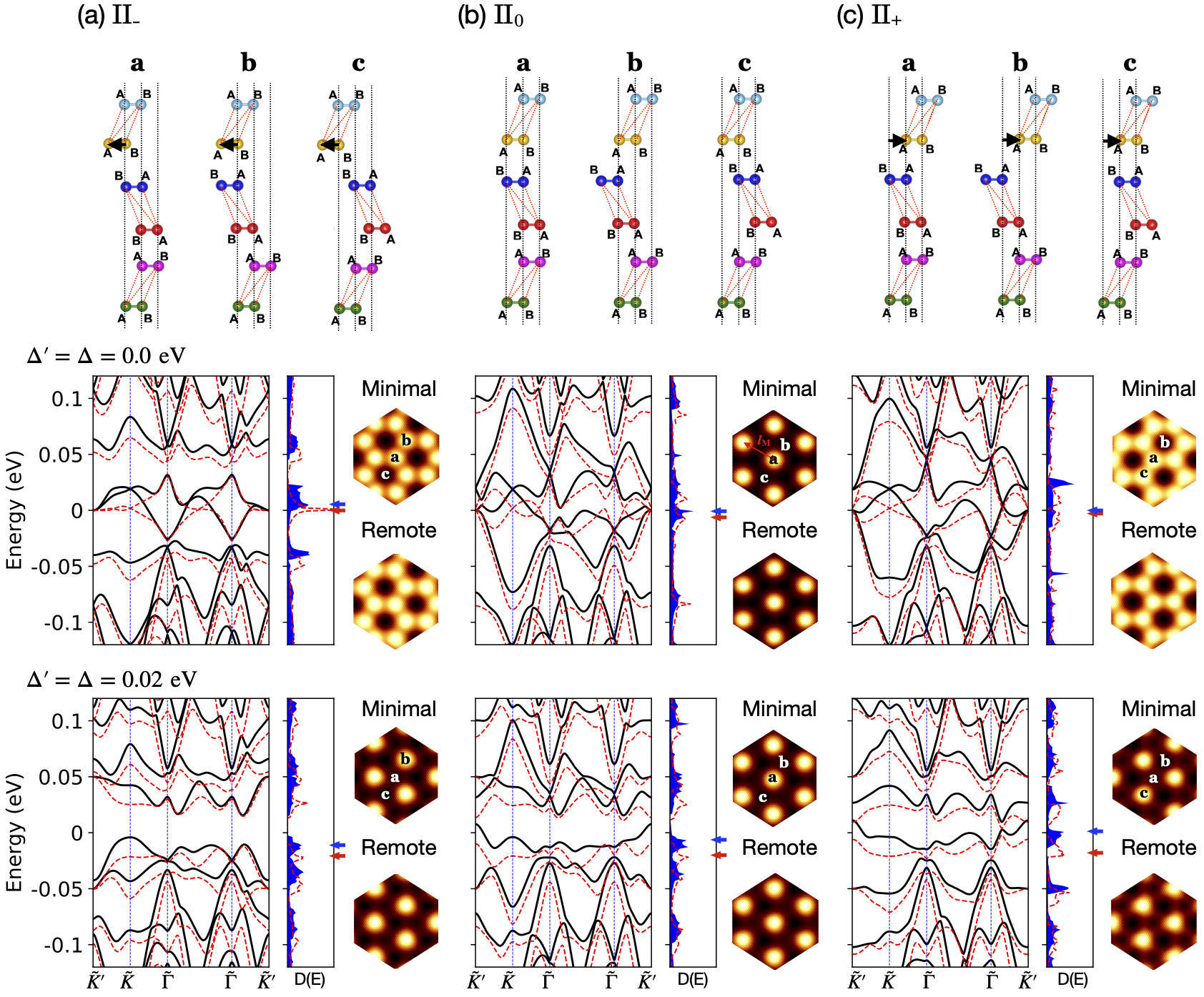}
\end{center}
\caption{
(Color online) Same as Fig.~\ref{fig4} for type II.
% Schematic diagrams for the atomic configurations of local commensurate stackings at {\bf a}, {\bf b}, and {\bf c} on the upper row, the band structures and the corresponding DOS for the minimal (red dotted lines) and for the remote hopping model (black solid lines), and their LDOS in the hexagonal frames when $\Delta' = \Delta = 0.0$ eV on the middle row, and when $\Delta' = \Delta = 0.02$ eV on the lower row for (a) type II$_-$, (b) type II$_0$, (c) type II$_+$.
}\label{fig5}
\end{figure*}
The $2\times2$ matrices $\bar{V}_i = V_i \mathbb{1}$ ($i = 1, \cdots, 6$) on the diagonal 
are the potential energies on the $i^{th}$ layer which can be expressed in terms of the interlayer potential differences within a 
bilayer $\Delta$ and the potential differences between bilayers $\Delta'$. 
When we assume that the potential $V_i$ drops monotonically at the same rate
neglecting screening effects, the $V_i$ values satisfy 
$V_1 = 3\Delta/2 +\Delta'$, $V_2 = \Delta/2 + \Delta'$, 
$V_3 = \Delta/2$, $V_4 = -V_3$, $V_5 = -V_2$, and $V_6 = -V_1$.
The interlayer tunneling terms within each bilayer are given by 2 $\times$ 2 matrix $t^{\pm}_s$ (where $s =$ AB or BA) consisting of 
(1) the vertical hopping term in the Bernal-stacked bilayer $t_1 = 0.361$~eV originating from $t_1(\sigma_x - i m_s \sigma_y)/2$ for either AB- or BA-stacked bilayer, 
and (2) the two additional hopping terms, $v_3$ and $v_4$, responsible for the trigonal warping and electron-hole asymmetry, respectively, as proposed in Ref.~\cite{Jung2014a}, where $v_i = \sqrt{3} \vert t_i \vert a/ 2\hbar$, $t_3 = 0.283$~eV and $t_4 = 0.183$~eV. %~\cite{Kuzmenko2009, Wong2015}. 
The explicit form of the matrix for an AB-stacked bilayer $t^{\pm}_{AB}$, as an example, is given as follows,
\begin{equation}
t^{\pm}_{AB} = \left(\begin{array}{cc}
-v_4 \pi^{\pm \dagger} & -v_3 \pi^{\pm}\\
t_1 & -v_4 \pi^{\pm \dagger} \\
\end{array}\right), 
\end{equation}
resulting in $t^{\pm}_{BA} = t^{\pm \dagger}_{AB}$. Here, $\pi^{\pm}$ represents $p_x + i p_y$ with an implicit phase of $\pm \theta/2$ due to the twist. In the minimal model the remote hopping terms are set to zero, i.e. $\delta = v_3 = v_4 = 0$.

\begin{figure*}
\begin{center}
\includegraphics[width=0.98\textwidth]{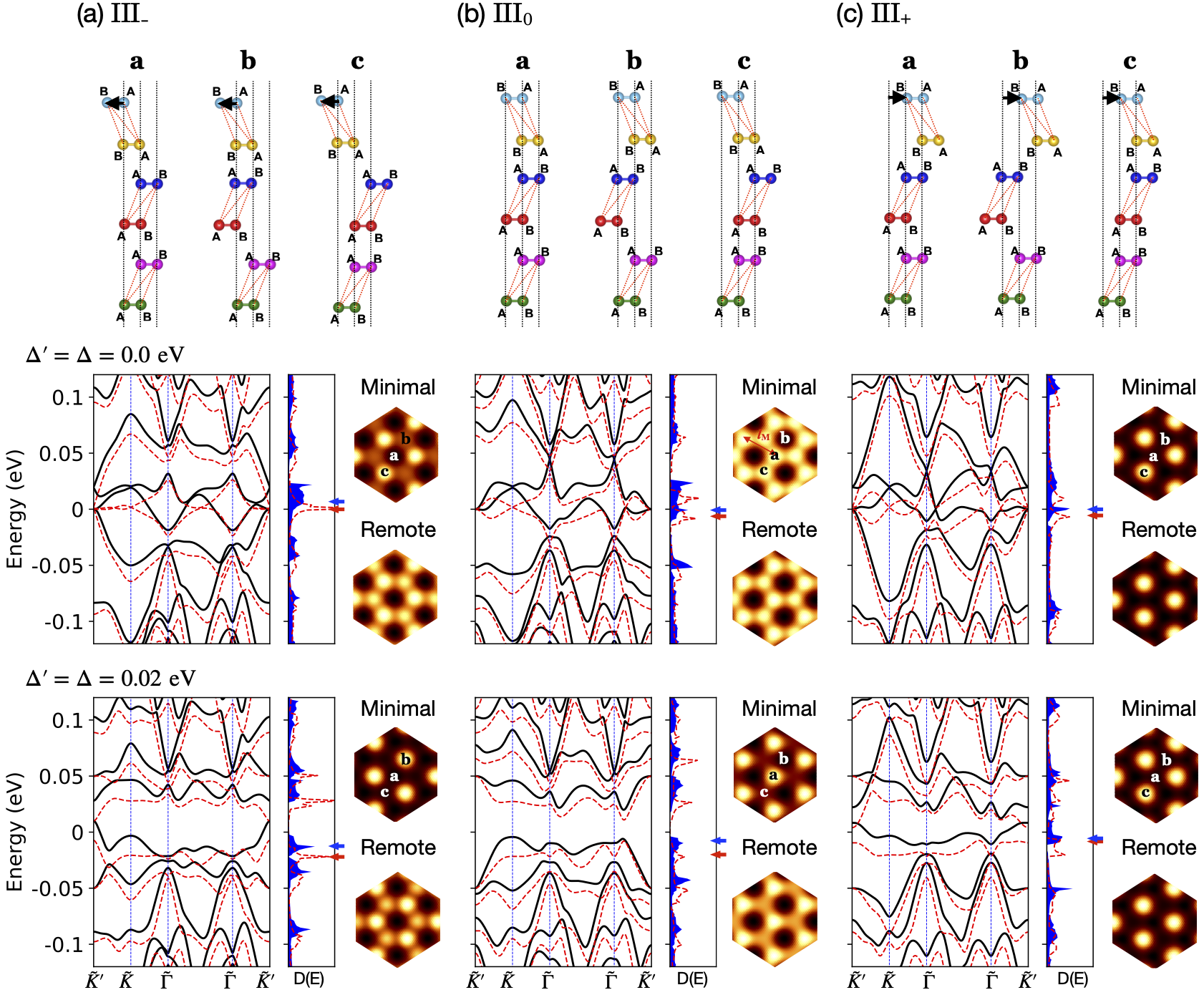}
\end{center}
\caption{
(Color online) Same as Fig.~\ref{fig4} for type III.
%Schematic diagrams for the atomic configurations of local commensurate stackings at {\bf a}, {\bf b}, and {\bf c} on the upper row, the band structures and the corresponding DOS for the minimal (red dotted lines) and for the remote hopping model (black solid lines), and their LDOS in the hexagonal frames when $\Delta' = \Delta = 0.0$ eV on the middle row, and when $\Delta' = \Delta = 0.02$ eV on the lower row for (a) type III$_-$, (b) type III$_0$, (c) type III$_+$.
}\label{fig6}
\end{figure*}

The interlayer tunneling at the twisted interface is captured by~\cite{Bistritzer2010b, Jung2014}
\begin{equation}
T_k(\boldsymbol{r}) = \sum_{j=0, \pm} e^{-i m_k \boldsymbol{q}_j \cdot \boldsymbol{r}} T^j_{ll'}.
\end{equation}
where $m_k = (-1)^k$, and $\boldsymbol{q}_0$, $\boldsymbol{q}_\pm$ are given as $\boldsymbol{q}_0 = \theta k_D (0, -1)$, $ \boldsymbol{q}_\pm = \theta k_D (\pm \sqrt{3}/2, 1/2)$ in the small angle approximation.
%if the twist angle $\theta$ is assumed to be small enough. 
Here, $k_D = 4\pi/3 a$ is equal to the length of a side of the Brillouin zone of a monolayer graphene where $a = 2.461 \textrm{\AA}$.

The interlayer tunneling at moire interfaces can be expressed in terms of the horizontal displacement vector $\bm{d}$ which measures the horizontal direction and distance between two atoms on the same sublattice -- one atom from the upper layer and the other from the lower layer. The coplanar displacement between two atoms of the same sublattice from the different layers, for example, is given as $\bm{d} = (d_x, d_y) = (0, 0)$ for AA$^{\prime}$ and BB$^{\prime}$ stackings, $(0, +a/\sqrt{3})$ for AB$^{\prime}$ stacking, and $(0, -a/\sqrt{3})$ for BA$^{\prime}$ stacking at the moire interfaces. Then, the interlayer tunneling is explicitly expressed as  $T_{ll’} = \omega_{ll'} \exp(ij \bm{G}_j \cdot \bm{d})$ for $j=0, \pm 1$ corresponding to $\bm{G}_0 = (0, 0)$ and $\bm{G}_\pm = k_D (-3/2, \pm \sqrt{3}/2)$ with sublattice indices $l$ and $l'$, yielding

\begin{equation}
T^0 = \left(\begin{array}{cc}
\omega' & \omega\\
\omega & \omega' \\
\end{array}\right), \hspace{0.3 cm} T^{\pm} = \left(\begin{array}{cc}
\omega' & \omega e^{\mp i 2\pi /3}\\
\omega e^{\pm i 2\pi /3} & \omega' \\
\end{array}\right),
\end{equation}
for AA stacking, for example. Here we use $\omega' = \omega_{AA'} = \omega_{BB'} = 0.0939$~eV 
for the intra-sublattice tunneling energy and $\omega = \omega_{AB'} = \omega_{BA'} = 0.12$~eV 
for the inter-sublattice tunneling at zero pressure
when we capture the out-of-plane lattice relaxation effects 
parametrized through the polynomial
%resulting from the exact exchange and random phase approximation (EXX+RPA) and local denisty approximation (LDA) for the interlayer tunnelings 
\begin{equation}
\omega' = C_2 \omega^2 + C_1 \omega + C_0 \label{eqome}
\end{equation}
where $C_2 = -0.5506$ eV$^{-1}$, $C_1 = 1.036$, and $C_0 = -0.02245$ eV as reported in Ref.~\cite{Chebrolu2019}. 
Otherwise, we can use equal $ \omega' =  \omega = 0.12$~eV when we do not have the lattice corrugations in the $z$-direction. 
%
%which are obtained by averaging all of possible tunneling components over the whole configurations and equilibrating the distance through the EXX+RPA \cite{Leconte2017}.
%

\begin{figure*}
\begin{center}
\includegraphics[width=1.0\textwidth]{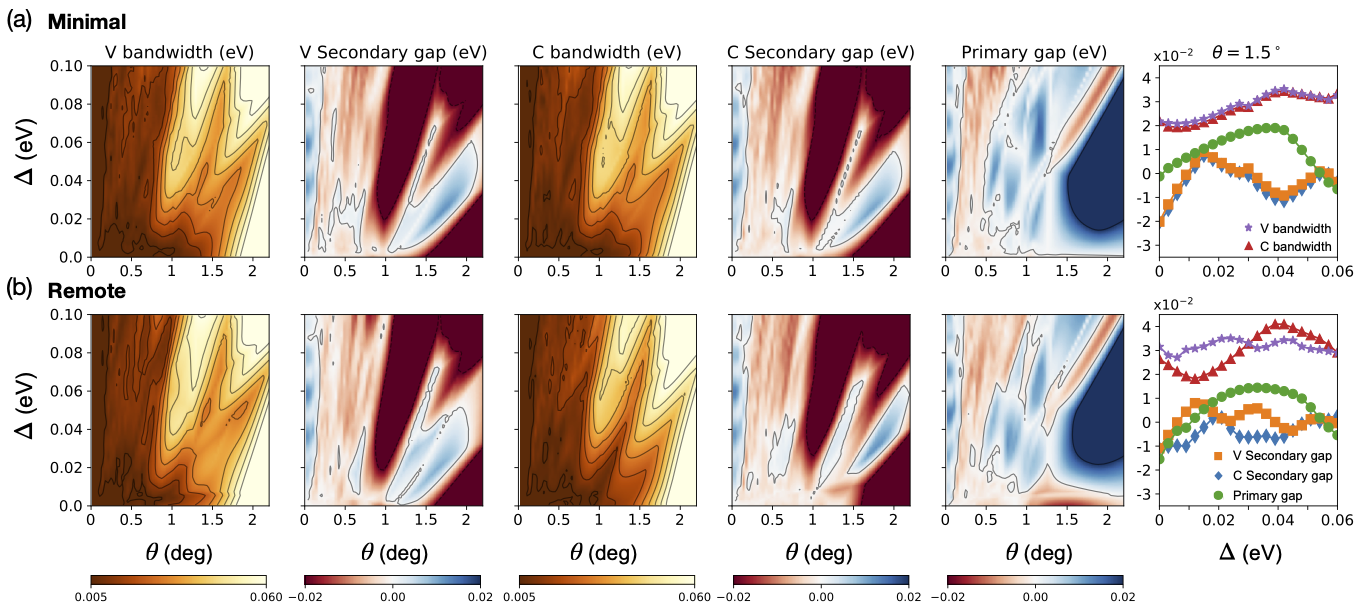}
\end{center}
\caption{
(Color online) Bandwidth, secondary bandgap, and primary bandgap of the lowest-energy valence (V) and conduction (C) bands of type I$_0$ [see Fig.~\ref{fig4} (b) for the atomic configurations of the local commensurate stackings] for (a) the minimal and (b) the remote hopping model as a function of the displacement field and the twist angle. Here we set $\Delta' = \Delta$ where $\Delta~(\Delta')$ represents the potential difference within (between) bilayers. The solid lines in the first and the third columns represent contours of the bandwidth of the lowest-energy valence and the conduction bands. The solid lines in the second, fourth and fifth columns represent the contour at the zero gaps. The cross-section at $\theta = 1.5^\circ$ is presented on the rightmost column with (purple) stars for the V bandwidth, (red) triangles for the C bandwidth, (orange) squares for the V secondary gap, (blue) diamonds for the C secondary gap, and (green) circles for the primary gap. 
}\label{fig7}
\end{figure*}

\begin{figure*}
\begin{center}
\includegraphics[width=1\textwidth]{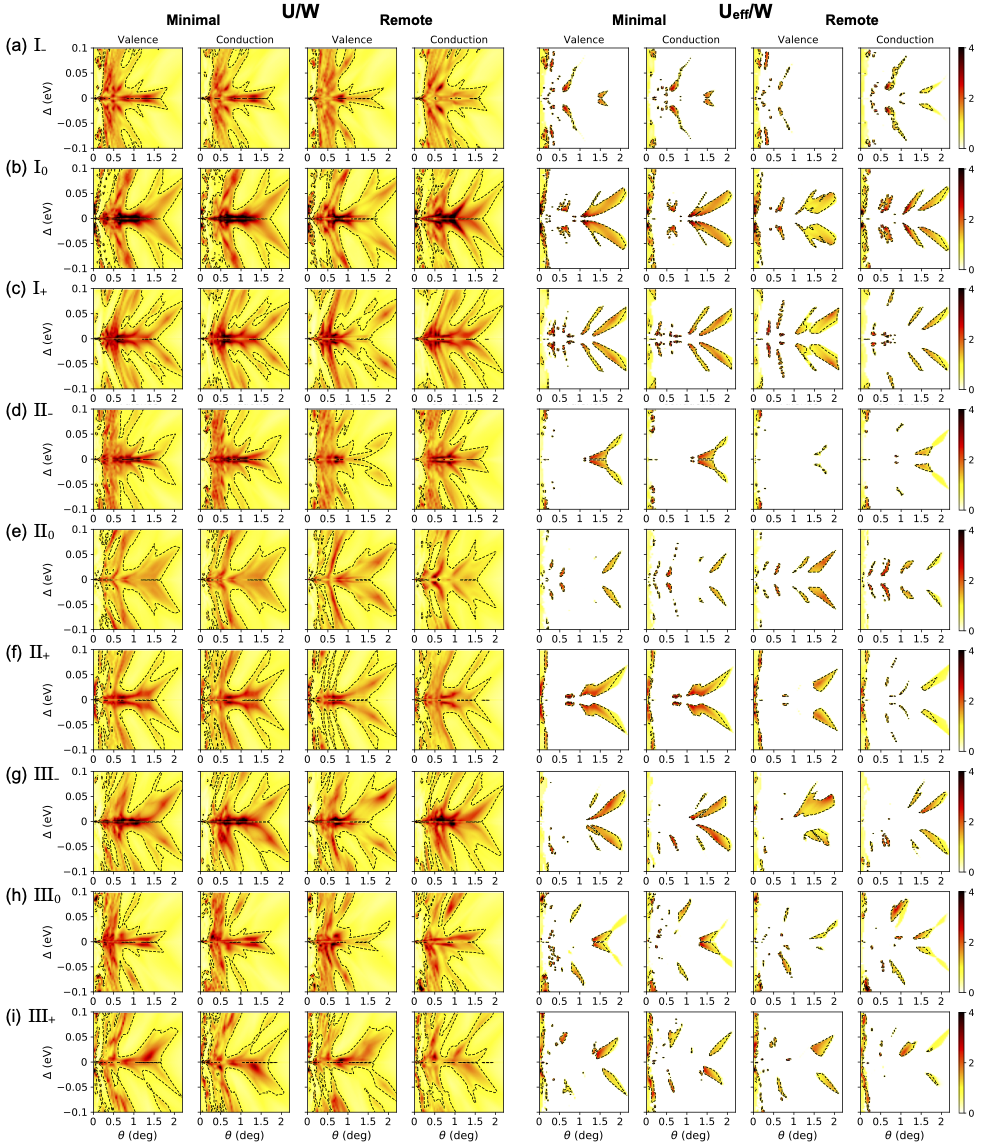}
\end{center}
\caption{
(Color online) The ratio of the bare Coulomb interaction to the bandwidth $U/W$ and the ratio of the effective Coulomb interaction to the bandwidth $U_\textrm{eff}/W$ [see Eq. (\ref{Ueff})] for the minimal and the remote hopping model as a function of the interlayer potential difference $\Delta$ and twist angle $\theta$ of the lowest-energy valence and conduction bands of type I$_{(-, 0, +)}$ in (a), (b), (c), type II$_{(-, 0, +)}$ in (d), (e), (f), and type III$_{(-, 0, +)}$. We set $\Delta' = \Delta$ where $\Delta~(\Delta')$ represents the potential difference within (between) bilayers.
}\label{fig8}
\end{figure*}

In Fig.~\ref{fig1}(e) we show schematic diagrams for four cases when the interlayer coupling $\omega$ at twisted interfaces is considered as a perturbation. We have three parabolic bands stemming from each bilayer as shown in the first panel where there is no interlayer coupling at moire interfaces. If we have a small interlayer coupling, for example, $\omega = 0.01$ eV, the one parabolic bands at $\tilde{K}'$ and the one at $\tilde{K}$ start hybridizing as denoted by $\Delta_c$ as seen in the second panel. $\Delta_c$ increases as the twist angle grows such that $\Delta_c \approx 7.3$ meV, $7.9$ meV, $8.9$ meV at $\theta = 1.0^\circ$, $1.5^\circ$, $2.5^\circ$, respectively for $\omega = 0.01$ eV. As shown in the third case, an external electric field causes a band offset $2\Delta' + \Delta$ between the two parabolic bands at $\tilde{K}'$ as well as the intrinsic gaps $\Delta$ in all three parabolic bands without the interlayer coupling at moire interfaces. 
In the fourth panel where we turn on interlayer coupling under the electric field $\Delta$, the strength of hybridization $\Delta_c'$ does not increase monotonically with respect to the interlayer coupling unlike $\Delta_c$ because of the another variable $\Delta$ in this case, and has values $\approx 3.3$ meV, $5.0$ meV, $3.9$ meV at $\theta = 1.0^\circ$, $1.5^\circ$, $2.5^\circ$, respectively. 

%--------------------------------------------------
\section{Electronic properties}\label{R1}
\subsection{Lattice relaxation and electric field}
% \subsection{Band dispersion}
% Fig. 2 (a), (b)
The electronic band structures for the 9 possible cases as a starting stacking due to (1) AB and BA alignments of the Bernal-stacked bilayer, and (2) translations of the top bilayer with respect to the bottom bilayer by $\bm{\tau}_0$, or $\bm{\tau}_\pm$ denoted as I$_{(0, +, -)}$, II$_{(0, +, -)}$, and III$_{(0, +, -)}$ are quite distinct from each other. Here we first consider the type I$_0$ and investigate the effects of the lattice relaxation and the electric field on the electronic band structures. See the Supplementary information for the analog of Fig.~\ref{fig2} for the other types. We present in Fig.~\ref{fig2} the energy dispersions and density of states (DOS) of the rigid lattice case ($\omega = \omega'$) and the case of the out-of-plane relaxed lattice ($\omega \neq \omega'$) for the minimal model in Figs.~\ref{fig2} (a) and \ref{fig2} (b), respectively, for the two twist angles, $\theta = 2 ^\circ$ and $1.5^\circ$, and for the two cases of interlayer potential differences, $\Delta' = \Delta = 0.0$ eV and $\Delta' = \Delta = 0.02$ eV.  
In the rigid lattice for zero electric field, at $\theta = 2^\circ$, the energy dispersions are similar in appearance such that the two parabolic bands located at $\tilde{K}$ and $\tilde{K}'$ are hybridized in common in all the 9 cases. It is noteworthy that the other parabolic band dispersion at $\tilde{K}'$ occurs differently according to the starting stackings and sliding types (not shown). It opens a large gap $\vert E \vert \gtrsim$ 0.05 eV at $\tilde{K}'$ in type I$_-$, II$_-$, and III$_0$, while it remains parabolic as seen in type I$_0$ (Fig.~\ref{fig2}), II$_0$, and III$_+$, or has a form of slightly linear bands in types I$_+$, II$_+$, and III$_-$.
At $\theta = 1.5 ^\circ$, the band dispersions are quite different for each type, however, they have a common feature that they have a gap at the $\tilde{\Gamma}$ point of the size of $E_{gap, \tilde{\Gamma}} \gtrsim$~0.05~eV regardless of the type. To be specific, for type I$_0$ as shown in Fig.~\ref{fig2}, the two parabolic bands make nearly flat bands with a gap-opening at $\tilde{\Gamma}$. 

It has been reported that a corrugation effect induced by out-of-plane lattice relaxation, which was observed in both experiments \cite{Cao2016, Cao2018a, Cao2018b} and theoretical works \cite{Uchida2014, VanWijk2015} causes flat bands to be isolated from adjacent dispersing bands \cite{Koshino2018}.
The changes in the electronic properties due to the lattice relaxation are captured
through unequal interlayer tunneling strengths at the moire interface, $\omega$ and $\omega'$ such that it follows the relation of Eq.~(\ref{eqome}). The effects appear remarkably in type I$_0$, II$_-$, III$_-$
lifting the degeneracies between the lowest-energy and the next higher-energy bands
whereas there are no qualitative changes in the electronic properties in the other types. 

On the other hand, the finite electric field $\Delta = 0.02$~eV opens a primary gap $\delta_p$ which is defined as a gap between the lowest-energy electron and the hole bands as indicated by the red shaded region on the right panel of lower row in Fig.~\ref{fig2} (b) for all the 9 cases for the relaxed lattice. The secondary gap $\delta_s$ which is defined as a gap between the lowest-energy electron (hole) band and the next higher-energy electron (hole) band as indicated by the blue shaded regions on the right panel of lower row in Fig.~\ref{fig2} (b) barely opens only in type I$_0$, I$_+$, II$_+$, III$_-$, and III$_+$ for the relaxed lattice.

% Fig. 2 (c), (d), (e), (f) 
In the first row of Figs.~\ref{fig2}(c) and \ref{fig2}(d), we show the bandwidths of the lowest-energy electron and hole bands of the type I$_0$ as a function of twist angle for the rigid and relaxed lattice cases, respectively. Due to the gap-opening at $\tilde{\Gamma}$, the bandwidth increases near $\theta = 1.5^\circ$ as the twist angle increases. 
The bandwidths evolve with twist angle similarly for different interlayer coupling strengths $\omega = 0.15$, $0.18$~eV but are shifted approximately linearly with $\omega$\cite{Chittari2018, Chebrolu2019}. The different interlayer tunneling can be achieved by applying pressure in the $z$-direction. We defer the detailed discussions on the effect of pressure to the next subsection. Interestingly, the t3BG system does not have a local minimum in bandwidth at the twist angle $\theta = 1.5^\circ$ where t3G has its first magic-angle\cite{Khalaf2019, Carr2019a, Li2019, tb_tTG,lei2020mirror, hao2020electric, Dumitru2021, Shin2021}, while t2BG has the local minimum in bandwidth at the magic-angle $\theta = 1.06^\circ$ which is also the magic-angle of t2G \cite{Chittari2018}. 
Figs.~\ref{fig2}~(e) and (f) give us the bandwidth as a function of $\omega$. The bandwidth appears to be smaller than 20~meV at $\theta = 1.5^\circ$ in the range of $\omega \gtrsim$~0.12~eV. This behavior repeats at $\theta = 1.0^\circ$ but in the range of $\omega \gtrsim$~0.09~eV and at $\theta = 2.0^\circ$ in the range of $\omega \gtrsim$~0.18~eV. The out-of-plane lattice relaxation makes the graph of bandwidth smoother but the overall physics does not change. 

Similar to t3G where the electronic structure can be decomposed into a pair of t2G at different angles with an additional decoupled monolayer graphene~\cite{Khalaf2019}, we find that the electronic structure of t3BG can be approximately described by a pair of t2BG at different angles with an additional decoupled bilayer graphene at low energies near $\tilde{K}$ for I$_0$ without an external field. Importantly, there are two parabolic bands at $\tilde{K}^\prime$ which are exactly overlaid indicated by the dotted purple lines, and one parabolic band at $\tilde{K}$ indicated by the dotted yellow lines in the first row in Figs.~\ref{fig2} (a) and \ref{fig2}(b). One of the two parabolic bands at $\tilde{K}^\prime$ corresponds to the decoupled bilayer graphene, and its effective mass remains the same irrespective of the twist angle under the zero electric field. For a finite electric field, the two parabolic bands at $\tilde{K}^\prime$ start hybridizing with each other and the effective masses of them change as indicated by the dotted purple, pink lines in the second row in Figs.~\ref{fig2} (a) and \ref{fig2}(b). 

%-----------------------------------------------------------------------------------------------------%
%
\subsection{Under pressure}
% Fig. 3
In Fig.~\ref{fig3} we present an extended perspective of the electronic band structure of I$_0$ under pressure for the relaxed lattice case such as the bandwidth of the lowest-energy electron and hole bands, the primary gap $\delta_p$, and the secondary gaps $\delta_s$. See the Supplementary information for a more complete survey. Based on the earlier studies~\cite{Jung2014, Leconte2017, Yankowitz2018, Chebrolu2019}, here we introduce the same strategy used in Ref.~\cite{Chebrolu2019} given by the two relations, between the external pressure and the interlayer distance, and between the interlayer tunneling $\omega$ and the interlayer distance, leading to a second-order polynomial as follows.

\begin{equation}
P = A_1 \omega^2 + A_2 \omega + A_3,
\end{equation}
where we use the fitting parameters $A_1 = 324.7$ GPa/eV$^2$, $A_2 = -35.47$ GPa/eV and $A_3 = -0.4671$ GPa obtained from EXX+RPA. The zero pressure (P = 0 GPa) is indicated by a black dashed line. The contour for the positive (negative) level is drawn by the solid (dashed) lines. We juxtapose two $y$-axes together -- one for the pressure in the unit of GPa on the left side, and the other for the interlayer tunneling strength $\omega$ on the right side.
We get electronic properties under a certain condition of external pressure such that there appears a linear contour with an approximate relation between the pressure and the twist angle such as $P \sim 3.3 \theta$ for both the minimal and remote hopping models for all the 9 cases as shown in Figs.~\ref{fig3}(a) and \ref{fig3}(b).
% $P \sim 3.266666 \theta$
Type II$_-$ and III$_0$ have finite secondary gaps even under the zero electric field, while others have finite secondary gaps only when the system is gated by an electric field. The result with the a finite interlayer potential difference $\Delta' = \Delta = 0.02$~eV for type I$_0$ is shown in Figs.~\ref{fig3}(c) and \ref{fig3}(d) as an example, and it has the positive secondary and primary gaps in the range of $\theta \gtrsim 0.5^\circ$ along the linear contour, while the primary and secondary gaps are closed in the entire parameter space under the zero interlayer potential difference. Furthermore, the finite interlayer potential difference tends to enlarge the size of the primary gap in both minimal and remote cases. The cross-section at $\theta = 1.5^\circ$ is illustrated on the rightmost column. In the minimal model without the displacement field the bandwidth (secondary gap) of the conduction is symmetric to that of the valence band regardless of the pressure. Upon turning on the remote hopping terms, the symmetry in bandwidth (secondary gap) between the conduction band and the valence band starts breaking as seen in the rightmost column in Figs.~\ref{fig3} (b) and \ref{fig3}(d). In contrast, the displacement field preserves the analogous features between the conduction and valence bands but turns the monotonically decreasing (increasing) bandwidth (secondary gap) into a non-monotonic function as the pressure increases as shown in the rightmost column in Fig.~\ref{fig3} (c).

\begin{figure*}
\begin{center}
\includegraphics[width=0.9\textwidth]{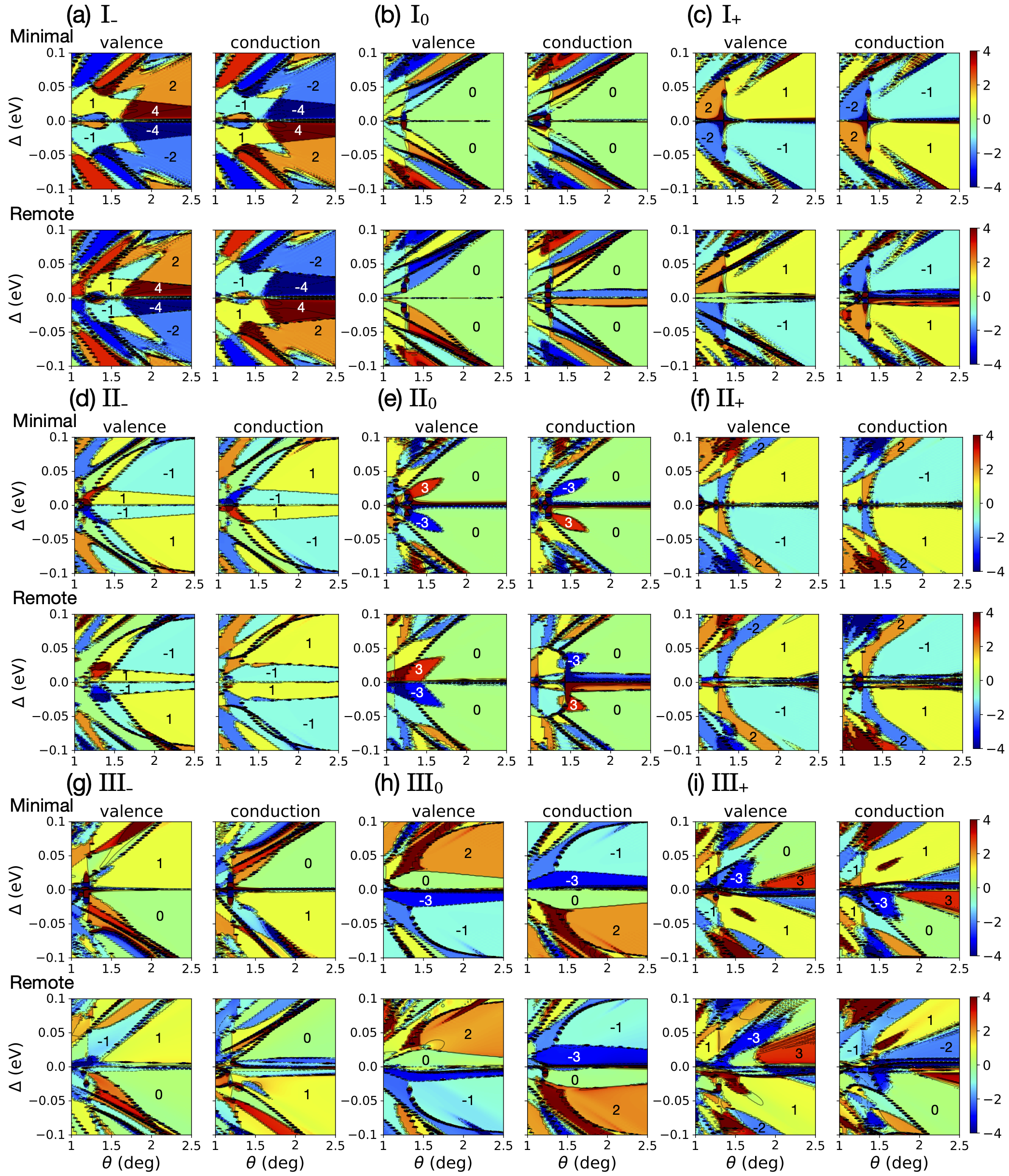}\\
\end{center}
\caption{
(Color online) Topological phase diagrams for the $K$-valley Chern numbers of the lowest-energy valence and conduction bands of t3BG of type I$_{(-, 0, +)}$ in (a), (b), (c), type II$_{(-, 0, +)}$ in (d), (e), (f), and type III$_{(-, 0, +)}$ in (g), (h), (i) in the minimal (upper row) and the remote (lower row) model as a function of the interlayer potential difference $\Delta$ and the twist angle. Here we use the condition $\Delta' = \Delta$ where $\Delta~(\Delta')$ represents the gated potential difference within (between) bilayers.
}\label{fig9}
\end{figure*}
%-----------------------------------------------------------------------------------------------------%
\subsection{Local density of states}
% Fig. 4, 5, 6
Each of the 9 stacking geometries shown in Figs.~\ref{fig1} (a)-(c) gives moire patterns consisting of three different local commensurate stackings, {\bf a}, {\bf b}, and {\bf c} as shown in Fig.~\ref{fig1} (d) with the three local commensurate stackings in the moire patterns for I$_0$ as an example. In principle, one can get the same band structure no matter which stackings among these three local commensurate stackings we take as the starting stacking. We listed the atomic configurations of the local commensurate stackings for type I$_{(-, 0, +)}$ in the first row in Fig.~\ref{fig4}, for type II$_{(-, 0, +)}$ in Fig.~\ref{fig5}, and for type III$_{(-, 0, +)}$ in Fig.~\ref{fig6}. Throughout Figs.~\ref{fig4}-\ref{fig6}, the twist angle is set to $\theta = 1.5^\circ$, the energy dispersions and DOS for the minimal model are indicated with red dashed lines, and the ones with the remote hopping terms are in black solid lines for the energy band and blue-filled areas for the DOS. 

The electronic features are quite distinct for the 9 different geometries {I, II, III}$_{(-, 0, +)}$ as shown in Figs.~\ref{fig4}-\ref{fig6}. It is common across all types of stackings that the minimal model has particle-hole symmetric band, and the introduction of the remote hopping terms breaks the electron-hole symmetry especially of the lowest-energy bands resulting in DOS broadening. 
%to break the symmetry in the electronic band structures, leading to biased van Hove singularities to one of the lowest-energy band or brodened DOS.
%
We show the band structures under a finite displacement field $\Delta' = \Delta = 0.02$ eV and it is noteworthy that all the 9 cases have a finite primary gap $\delta_p$ and some of cases have finite secondary gaps $\delta_s$ as well, leading to the isolated lowest-energy bands.

% LDOS
We show the local density of states (LDOS) at the van Hove singularities denoted by the red arrow for the minimal and the blue arrow for the remote hopping terms model. We found in common for all the 9 cases that the local-AA stacking regions at the moire interface have relatively high LDOS concentrations. For instance, among the local commensurate stackings of type I$_-$, the stacking {\bf c} has the local-AA configurations at both moire interfaces, yielding the highly concentrated LDOS, unlike the stackings {\bf a} and {\bf b}. In type I$_0$, on the other hand, the local commensurate stackings {\bf b} and {\bf c} have the local-AA configurations at one of the two moire interfaces but {\bf a} does not have the local-AA stackings at both interfaces. Thus, the intensity of LDOS at the stackings {\bf b} and {\bf c} surpasses that of the stacking {\bf a} for a moire length of $l_M = 9.4$~nm.
%is in close agreement with the analytic value $a/2 \sin(\theta/2)$.}
%-----------------------------------------------------------------------------------------------------%

\subsection{Bandwidths and bandgaps}
% Fig. 7
In the following, 
%a continuation of having studied the effect of the displacement field on the electronic structures, here 
we explore the evolution of the bandwidths, secondary, and primary bandgaps in the parameter space of the displacement field, and the twist angle for I$_0$ taking Fig.~\ref{fig7} as an example. See the supplementary information for other stackings.
%%%%%% GJ %%%%%%
All 9 starting stacking cases have in common that the valence and conduction bandwidths are not suppressed around $\theta = 1.5^\circ$ but the secondary gaps $\delta_s$ become open at several large or small islands around this angle for finite displacement fields. Besides, the primary gap $\delta_p$ is open in a large range of the parameter space, 
a fact that will facilitate the band isolation. 
The rightmost column represents cross-sections at $\theta = 1.5^\circ$ for the bandwidths and bandgaps. In these plots the difference between both minimal and remote hopping models becomes more apparent that the electron-hole symmetry of the bands in the minimal model is preserved
except for type III$_+$ where the electron-hole asymmetry with respect to electric field is already broken in the minimal model. 
More specifically, the type I$_0$ in Fig.~\ref{fig7} has finite secondary gaps at $\theta \lesssim 0.2^\circ$ for $\Delta > 0$, at several small islands $0.2^\circ \lesssim \theta \lesssim 0.7^\circ$ in the range of $\Delta \lesssim 0.04$ eV, and at larger islands $1^\circ \lesssim \theta \lesssim 2^\circ$ in the range of $\Delta \lesssim 0.06$ eV.
%-----------------------------------------------------------------------------------------------------%

\begin{figure*}
\begin{center}
\includegraphics[width=0.95\textwidth]{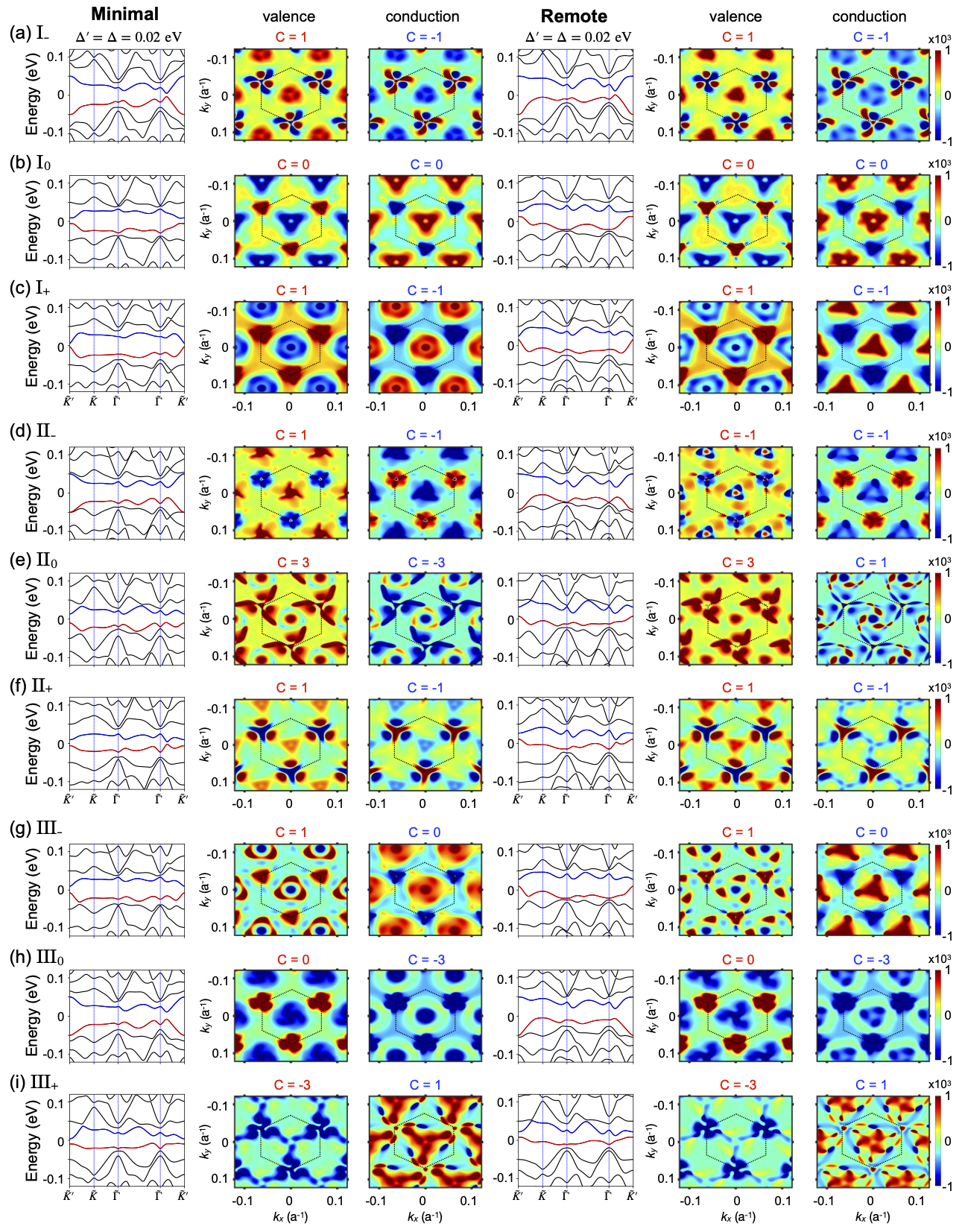}\\
\end{center}
\caption{
(Color online) Band structures at $\theta = 1.5^\circ$ and the corresponding Berry curvatures for the lowest-energy valence and conduction bands with the $K$-valley Chern number [see Fig.~\ref{fig9}] on top of them for type I$_{(-, 0, +)}$ in (a), (b), (c), type II$_{(-, 0, +)}$ in (d), (e), (f), and type III$_{(-, 0, +)}$ in (g), (h), (i). Here we set $\Delta' = \Delta = 0.02$~eV for both the minimal model and the remote hopping model where $\Delta~(\Delta')$ represents the potential differences within (between) bilayers. 
}\label{fig10}
\end{figure*}

%-----------------------------------------------------------------------------------------------------%
\section{Effective Coulomb interactions}\label{R2}
We evaluate the relative strength of the Coulomb interactions with partial screening due to the metallicity of the bands as in Ref.~\cite{Chebrolu2019}:
\begin{equation}
U_\textrm{eff} = U \exp{(-l_M/ \lambda_D)},
\label{Ueff}
\end{equation}
where $U$ represents the bare Coulomb interaction which is defined as $\frac{e^2}{4 \pi \epsilon_r \epsilon_0 l_M}$, the moire length $l_M \sim a/\theta$ when the twist angle $\theta$ is small enough, and $\lambda_D$ is the Debye length which can be expressed as $\lambda_D = 2 \epsilon_0 / [e^2 D(\delta_p, \delta_s)]$ with the two-dimensional DOS $D(\delta_p, \delta_s)$. Here, we define $D(\delta_p, \delta_s)$ as
\begin{equation}
D(\delta_p, \delta_s) = 4\frac{\vert \delta_p \vert u(-\delta_p) + \vert \delta_s \vert u(-\delta_s)}{W^2 A_M},
\end{equation}
where $u(x)$ is the Heaviside step function so that $D(\delta_p, \delta_s)$ is proportional to the ratio of overlapping bands to the bandwidth $|\delta_{p(s)}|/W$ if $\delta_p < 0$ ($\delta_s < 0$), $A_M$ is the area of a moire unit cell in real space given as $\sqrt{3} ~ l^2_M/2$. In the current work, we set the dielectric constant of graphene as $\epsilon_r = 4$ \cite{Jung2013}.

% Fig 8
We discuss the ratio of the bare $U$ and the effective Coulomb interaction $U_\textrm{eff}$ to the bandwidth $W$ as a function of the displacement field within bilayers $\Delta$ and the twist angle for both minimal and remote hopping models for the all 9 cases I$_{(-, 0, +)}$,  II$_{(-, 0, +)}$, and III$_{(-, 0, +)}$ shown in Fig.~\ref{fig8}. 
For all cases, the bare Coulomb interaction over the bandwidth has the value of $U/W \gtrsim 1$ in almost the entire parameter space and has the maximum values at around $0.2^\circ \lesssim \theta \lesssim 1.5^\circ$ for weak displacement fields  $ \vert \Delta \vert \lesssim 25$~meV. Apart from type I$_-$, II$_-$, III$_0$, the strongly correlated electron states extended like a long tail above $ \vert \Delta \vert \sim 25$~meV for both the minimal and remote models. On the other hand, if we take into account the screening effect due to the conducting electrons, namely the effective Coulomb interactions $U_\textrm{eff}$, the territory where the strongly correlated states appear is suppressed for both the minimal and remote hopping cases and remain finite only at small islands in the range of $1 ^\circ \lesssim \theta \lesssim 2 ^\circ$ when $\vert \Delta \vert \lesssim 0.05$~eV in all the cases. 
To be more specific, in type I$_0$, I$_+$, and III$_-$ relatively larger islands survive under the inclusion of the screening effects by the itinerant electrons in the valence band of the remote hopping model at $\theta = 1.5^\circ$ under a week displacement field $\vert \Delta \vert \lesssim 0.05$~eV. On the other hand, the strong effective Coulomb regions are barely found at around $\theta = 1.5^\circ$ in type I$_-$, I$_+$, II$_0$, II$_+$, and III$_+$ since it is cut off by the electrons in the first and second lowest-energy bands overlapping with each other.
It is also noteworthy in Fig.~\ref{fig8} that both $U/W$ and $U_\textrm{eff}/W$ are symmetric against $\Delta = 0$~eV line except the three types III$_{(0, -, +)}$ where the stacking configurations of the bilayers on top (BA) and bottom (AB) are different to each other.
%-----------------------------------------------------------------------------------------------------%
\section{Valley Chern numbers}\label{R3}
We now turn our attention to the topological nature of the bands in t3BG by evaluating the Chern number of the lowest-energy valence and conduction bands only for one valley. Here we calculate the Chern number of $n^{\rm th}$ energy band $C_n$ for the $+K$ valley which is defined as
\begin{equation}
C_n = \frac{1}{2 \pi} \int_\textrm{mBZ} d^2 \boldsymbol{k} ~ \Omega_n (\boldsymbol{k}),
\end{equation}
by integrating the Berry curvature $\Omega_n (\boldsymbol{k})$ within the moire Brillouin zone (mBZ) which is given as \cite{Xiao2010b}
\begin{equation}
\Omega_n (\boldsymbol{k}) = -2 \sum_{n' \neq n} \textrm{Im} \Bigg[ \frac{\langle n \vert \frac{\partial H}{\partial k_x} \vert n' \rangle \langle n' \vert \frac{\partial H}{\partial k_y} \vert n \rangle}{(E_{n'} - E_n)^2} \Bigg].
\label{Bcurv}
\end{equation}
% Fig 9
Here, $E_n$ represents an energy eigenvalue of the moire superlattice Bloch state $\vert n \rangle$. 

In Fig.~\ref{fig9} we show the $K$-valley Chern numbers of the lowest-energy valence and conduction bands for all 9 starting stackings without/with remote hopping terms. In all cases, the valence and conduction bands have well-defined Chern numbers 
%between -4 and 4 
in almost the entire parameter space of the displacement field within bilayers $\vert \Delta \vert \leq 0.1$ eV and the twist angle $1^\circ \leq \theta \leq 2.5^\circ$. The inclusion of the remote hopping terms barely changes the phase diagram but the borders between topologically distinct phases are slightly shifted. Apart from the three types III$_{(-, 0, +)}$ the $K$-valley Chern numbers of the valence and conduction bands tend to be the opposite to each other such that the sum of them is zero. 

For the I$_0$, I$_+$, and III$_-$ cases where the regions of strongly correlated states survive in a slightly larger area under the screening effects in the remote hopping model as discussed in the previous section, type I$_0$ has the topologically trivial valence and conduction bands near $\theta = 1.5^\circ$, while type I$_+$ has a topologically nontrivial phase with C$=+1~(-1)$ in the valence (conduction) band for $\Delta > 0$ [C$=-1~(+1)$ in the valence (conduction) band for $\Delta < 0$], and type III$_-$ with C$=+1~(0)$ in the valence (conduction) band for $\Delta > 0$ [C$=0~(+1)$ in the valence (conduction) band for $\Delta < 0$] in the areas of the correlated electrons.

% Fig 10
In Fig.~\ref{fig10} we present the band structures and the corresponding Berry curvatures $\Omega_n (\bm{k})$ for the lowest-energy electron and hole bands at $\theta = 1.5^\circ$ for all 9 cases I$_{(-, 0, +)}$, II$_{(-, 0, +)}$, and III$_{(-, 0, +)}$ under a certain positive finite value of interlayer difference field, $\Delta' = \Delta = 0.02$ eV for both the minimal and remote cases. We highlight the valence (conduction) band with the red (blue) line in the band structure figure and indicate the individual band topological invariant on top of each Berry curvature figure. The mBZ is indicated by a black solid hexagon. 
Thus one can see the effect of the remote hopping terms on the Berry curvatures. We see that the topological properties generally remain intact upon inclusion of the remote hopping terms except for the valence band of type II$_-$ and the conduction band of type II$_0$.

%-----------------------------------------------------------------------------------------------------%
\section{SUMMARY}\label{Summary}

We have presented the electronic structure analysis of 
alternatingly twisted triple Bernal-stacked bilayer graphene in search of new multilayer graphene devices that give rise to 
nearly flat lowest-energy Chern bands. Twisted systems made up of bilayer graphene are interesting because each layer can open a bandgap upon application of an electric field. Since they have distinct AB or BA stacking orientations we have considered 9 starting stackings by considering three different sliding vectors $\bm{\tau}_{0, \pm}$ for the top bilayer with respect to the bottom bilayer.

More specifically, we have first described the detailed electronic properties, such as bandwidths and bandgaps of the lowest-energy bands in a wide parameter space of the interlayer couplings strength $\omega$ varied by the pressure, the twist angle $\theta$, and the interlayer potential difference $\Delta$ in the perpendicular direction that can be introduced by an electric field. 
We have found that in all 9 starting stackings, the lowest-energy bandwidths can be easily suppressed to values of the order of $\sim$20~meV near $\theta \approx 1.5^\circ$ and even narrower bands can be obtained for smaller twist angles.
We have compared the band structures for the minimal model with the remote hopping model, and we have shown the DOS and LDOS map giving rise to relatively high densities at the local-AA stacking sites of the moire interfaces. 

Our numerical analysis shows the electric field and twist angles where strong electron-electron interactions can be expected from the ratio between the bandwidth and the screened Coulomb potentials. We have found that type I$_0$, I$_+$, and III$_-$ are more favorable than other stackings to find correlated phases in the vicinity of $\theta = 1.5^\circ$ under a finite interlayer potential difference of $\Delta \sim 20$~meV.

The valley Chern numbers phase diagrams for the lowest-energy electron and hole bands for the minimal and remote hopping models have been obtained for all 9 stacking configurations 
for a range of twist angles below 2.5$^{\circ}$ and $\left| \Delta \right| < 0.1$~eV
and the associated Berry curvatures in the momentum space at the twist angle of $\theta = 1.5^\circ$ and $\Delta~(\Delta') \approx 20$ meV. 
The well-defined valley Chern numbers in the minimal model are mostly robust upon inclusion of the remote hopping terms except for two cases, the valence band in type II$_-$ and the conduction band in type II$_0$, where the remote hopping terms lead to band-touching with higher-energy bands. 
We expect that our work will provide guidance in the search of nearly flat bands in twisted triple bilayer devices.

\begin{acknowledgments}
This work was supported by Samsung Science and Technology Foundation Grant No. SSTF-BA1802-06 (J.S.), Korean NRF through the Grants No. 2021R1A6A3A01087281 (J.S.), No. 2020R1A5A1016518 (B.L.C.), No. 2018R1A2B6007837 (Y.J. and H.M.), No. 2020R1A2C3009142 (J.J.), and Creative-Pioneering Researchers Program through Seoul National University (SNU) (Y.J. and H.M.). We acknowledge computational support from KISTI Grant No. KSC-2021-CRE-0389 and by the resources of Urban Big data and AI Institute (UBAI) at UOS. J.J also acknowledges support by the Korean Ministry of Land, Infrastructure and Transport (MOLIT) from the Innovative Talent Education Program for Smart Cities and the KREONET network infrastructure.
\end{acknowledgments}

\bibliography{t3BG}

%merlin.mbs apsrev4-1.bst 2010-07-25 4.21a (PWD, AO, DPC) hacked
%Control: key (0)
%Control: author (72) initials jnrlst
%Control: editor formatted (1) identically to author
%Control: production of article title (-1) disabled
%Control: page (0) single
%Control: year (1) truncated
%Control: production of eprint (0) enabled
\begin{thebibliography}{47}%
\makeatletter
\providecommand \@ifxundefined [1]{%
 \@ifx{#1\undefined}
}%
\providecommand \@ifnum [1]{%
 \ifnum #1\expandafter \@firstoftwo
 \else \expandafter \@secondoftwo
 \fi
}%
\providecommand \@ifx [1]{%
 \ifx #1\expandafter \@firstoftwo
 \else \expandafter \@secondoftwo
 \fi
}%
\providecommand \natexlab [1]{#1}%
\providecommand \enquote  [1]{``#1''}%
\providecommand \bibnamefont  [1]{#1}%
\providecommand \bibfnamefont [1]{#1}%
\providecommand \citenamefont [1]{#1}%
\providecommand \href@noop [0]{\@secondoftwo}%
\providecommand \href [0]{\begingroup \@sanitize@url \@href}%
\providecommand \@href[1]{\@@startlink{#1}\@@href}%
\providecommand \@@href[1]{\endgroup#1\@@endlink}%
\providecommand \@sanitize@url [0]{\catcode `\\12\catcode `\$12\catcode
  `\&12\catcode `\#12\catcode `\^12\catcode `\_12\catcode `\%12\relax}%
\providecommand \@@startlink[1]{}%
\providecommand \@@endlink[0]{}%
\providecommand \url  [0]{\begingroup\@sanitize@url \@url }%
\providecommand \@url [1]{\endgroup\@href {#1}{\urlprefix }}%
\providecommand \urlprefix  [0]{URL }%
\providecommand \Eprint [0]{\href }%
\providecommand \doibase [0]{http://dx.doi.org/}%
\providecommand \selectlanguage [0]{\@gobble}%
\providecommand \bibinfo  [0]{\@secondoftwo}%
\providecommand \bibfield  [0]{\@secondoftwo}%
\providecommand \translation [1]{[#1]}%
\providecommand \BibitemOpen [0]{}%
\providecommand \bibitemStop [0]{}%
\providecommand \bibitemNoStop [0]{.\EOS\space}%
\providecommand \EOS [0]{\spacefactor3000\relax}%
\providecommand \BibitemShut  [1]{\csname bibitem#1\endcsname}%
\let\auto@bib@innerbib\@empty
%</preamble>
\bibitem [{\citenamefont {Kim}\ \emph {et~al.}(2017)\citenamefont {Kim},
  \citenamefont {DaSilva}, \citenamefont {Huang}, \citenamefont {Fallahazad},
  \citenamefont {Larentis}, \citenamefont {Taniguchi}, \citenamefont
  {Watanabe}, \citenamefont {LeRoy}, \citenamefont {MacDonald},\ and\
  \citenamefont {Tutuc}}]{Kim2017}%
  \BibitemOpen
  \bibfield  {author} {\bibinfo {author} {\bibfnamefont {K.}~\bibnamefont
  {Kim}}, \bibinfo {author} {\bibfnamefont {A.}~\bibnamefont {DaSilva}},
  \bibinfo {author} {\bibfnamefont {S.}~\bibnamefont {Huang}}, \bibinfo
  {author} {\bibfnamefont {B.}~\bibnamefont {Fallahazad}}, \bibinfo {author}
  {\bibfnamefont {S.}~\bibnamefont {Larentis}}, \bibinfo {author}
  {\bibfnamefont {T.}~\bibnamefont {Taniguchi}}, \bibinfo {author}
  {\bibfnamefont {K.}~\bibnamefont {Watanabe}}, \bibinfo {author}
  {\bibfnamefont {B.~J.}\ \bibnamefont {LeRoy}}, \bibinfo {author}
  {\bibfnamefont {A.~H.}\ \bibnamefont {MacDonald}}, \ and\ \bibinfo {author}
  {\bibfnamefont {E.}~\bibnamefont {Tutuc}},\ }\href {\doibase
  10.1073/pnas.1620140114} {\bibfield  {journal} {\bibinfo  {journal} {PNAS}\
  }\textbf {\bibinfo {volume} {114}},\ \bibinfo {pages} {3364} (\bibinfo {year}
  {2017})}\BibitemShut {NoStop}%
\bibitem [{\citenamefont {Cao}\ \emph {et~al.}(2018{\natexlab{a}})\citenamefont
  {Cao}, \citenamefont {Fatemi}, \citenamefont {Demir}, \citenamefont {Fang},
  \citenamefont {Tomarken}, \citenamefont {Luo}, \citenamefont
  {Sanchez-Yamagishi}, \citenamefont {Watanabe}, \citenamefont {Taniguchi},
  \citenamefont {Kaxiras}, \citenamefont {Ashoori},\ and\ \citenamefont
  {Jarillo-Herrero}}]{Cao2018a}%
  \BibitemOpen
  \bibfield  {author} {\bibinfo {author} {\bibfnamefont {Y.}~\bibnamefont
  {Cao}}, \bibinfo {author} {\bibfnamefont {V.}~\bibnamefont {Fatemi}},
  \bibinfo {author} {\bibfnamefont {A.}~\bibnamefont {Demir}}, \bibinfo
  {author} {\bibfnamefont {S.}~\bibnamefont {Fang}}, \bibinfo {author}
  {\bibfnamefont {S.~L.}\ \bibnamefont {Tomarken}}, \bibinfo {author}
  {\bibfnamefont {J.~Y.}\ \bibnamefont {Luo}}, \bibinfo {author} {\bibfnamefont
  {J.~D.}\ \bibnamefont {Sanchez-Yamagishi}}, \bibinfo {author} {\bibfnamefont
  {K.}~\bibnamefont {Watanabe}}, \bibinfo {author} {\bibfnamefont
  {T.}~\bibnamefont {Taniguchi}}, \bibinfo {author} {\bibfnamefont
  {E.}~\bibnamefont {Kaxiras}}, \bibinfo {author} {\bibfnamefont {R.~C.}\
  \bibnamefont {Ashoori}}, \ and\ \bibinfo {author} {\bibfnamefont
  {P.}~\bibnamefont {Jarillo-Herrero}},\ }\href {\doibase 10.1038/nature26154}
  {\bibfield  {journal} {\bibinfo  {journal} {Nature}\ }\textbf {\bibinfo
  {volume} {556}},\ \bibinfo {pages} {80} (\bibinfo {year}
  {2018}{\natexlab{a}})}\BibitemShut {NoStop}%
\bibitem [{\citenamefont {Cao}\ \emph {et~al.}(2018{\natexlab{b}})\citenamefont
  {Cao}, \citenamefont {Fatemi}, \citenamefont {Fang}, \citenamefont
  {Watanabe}, \citenamefont {Taniguchi}, \citenamefont {Kaxiras},\ and\
  \citenamefont {Jarillo-Herrero}}]{Cao2018b}%
  \BibitemOpen
  \bibfield  {author} {\bibinfo {author} {\bibfnamefont {Y.}~\bibnamefont
  {Cao}}, \bibinfo {author} {\bibfnamefont {V.}~\bibnamefont {Fatemi}},
  \bibinfo {author} {\bibfnamefont {S.}~\bibnamefont {Fang}}, \bibinfo {author}
  {\bibfnamefont {K.}~\bibnamefont {Watanabe}}, \bibinfo {author}
  {\bibfnamefont {T.}~\bibnamefont {Taniguchi}}, \bibinfo {author}
  {\bibfnamefont {E.}~\bibnamefont {Kaxiras}}, \ and\ \bibinfo {author}
  {\bibfnamefont {P.}~\bibnamefont {Jarillo-Herrero}},\ }\href {\doibase
  10.1038/nature26160} {\bibfield  {journal} {\bibinfo  {journal} {Nature}\
  }\textbf {\bibinfo {volume} {556}},\ \bibinfo {pages} {43} (\bibinfo {year}
  {2018}{\natexlab{b}})}\BibitemShut {NoStop}%
\bibitem [{\citenamefont {Yankowitz}\ \emph {et~al.}(2019)\citenamefont
  {Yankowitz}, \citenamefont {Chen}, \citenamefont {Polshyn}, \citenamefont
  {Zhang}, \citenamefont {Watanabe}, \citenamefont {Taniguchi}, \citenamefont
  {Graf}, \citenamefont {Young},\ and\ \citenamefont {Dean}}]{Yankowitz2018a}%
  \BibitemOpen
  \bibfield  {author} {\bibinfo {author} {\bibfnamefont {M.}~\bibnamefont
  {Yankowitz}}, \bibinfo {author} {\bibfnamefont {S.}~\bibnamefont {Chen}},
  \bibinfo {author} {\bibfnamefont {H.}~\bibnamefont {Polshyn}}, \bibinfo
  {author} {\bibfnamefont {Y.}~\bibnamefont {Zhang}}, \bibinfo {author}
  {\bibfnamefont {K.}~\bibnamefont {Watanabe}}, \bibinfo {author}
  {\bibfnamefont {T.}~\bibnamefont {Taniguchi}}, \bibinfo {author}
  {\bibfnamefont {D.}~\bibnamefont {Graf}}, \bibinfo {author} {\bibfnamefont
  {A.~F.}\ \bibnamefont {Young}}, \ and\ \bibinfo {author} {\bibfnamefont
  {C.~R.}\ \bibnamefont {Dean}},\ }\href {\doibase 10.1126/science.aav1910}
  {\bibfield  {journal} {\bibinfo  {journal} {Science}\ }\textbf {\bibinfo
  {volume} {363}},\ \bibinfo {pages} {1059} (\bibinfo {year}
  {2019})}\BibitemShut {NoStop}%
\bibitem [{\citenamefont {Cao}\ \emph {et~al.}(2020{\natexlab{a}})\citenamefont
  {Cao}, \citenamefont {Chowdhury}, \citenamefont {Rodan-Legrain},
  \citenamefont {Rubies-Bigorda}, \citenamefont {Watanabe}, \citenamefont
  {Taniguchi}, \citenamefont {Senthil},\ and\ \citenamefont
  {Jarillo-Herrero}}]{Cao2019}%
  \BibitemOpen
  \bibfield  {author} {\bibinfo {author} {\bibfnamefont {Y.}~\bibnamefont
  {Cao}}, \bibinfo {author} {\bibfnamefont {D.}~\bibnamefont {Chowdhury}},
  \bibinfo {author} {\bibfnamefont {D.}~\bibnamefont {Rodan-Legrain}}, \bibinfo
  {author} {\bibfnamefont {O.}~\bibnamefont {Rubies-Bigorda}}, \bibinfo
  {author} {\bibfnamefont {K.}~\bibnamefont {Watanabe}}, \bibinfo {author}
  {\bibfnamefont {T.}~\bibnamefont {Taniguchi}}, \bibinfo {author}
  {\bibfnamefont {T.}~\bibnamefont {Senthil}}, \ and\ \bibinfo {author}
  {\bibfnamefont {P.}~\bibnamefont {Jarillo-Herrero}},\ }\href {\doibase
  10.1103/PhysRevLett.124.076801} {\bibfield  {journal} {\bibinfo  {journal}
  {Phys. Rev. Lett.}\ }\textbf {\bibinfo {volume} {124}},\ \bibinfo {pages}
  {076801} (\bibinfo {year} {2020}{\natexlab{a}})}\BibitemShut {NoStop}%
\bibitem [{\citenamefont {Park}\ \emph
  {et~al.}(2021{\natexlab{a}})\citenamefont {Park}, \citenamefont {Cao},
  \citenamefont {Watanabe}, \citenamefont {Taniguchi},\ and\ \citenamefont
  {Jarillo-Herrero}}]{Park2021}%
  \BibitemOpen
  \bibfield  {author} {\bibinfo {author} {\bibfnamefont {J.~M.}\ \bibnamefont
  {Park}}, \bibinfo {author} {\bibfnamefont {Y.}~\bibnamefont {Cao}}, \bibinfo
  {author} {\bibfnamefont {K.}~\bibnamefont {Watanabe}}, \bibinfo {author}
  {\bibfnamefont {T.}~\bibnamefont {Taniguchi}}, \ and\ \bibinfo {author}
  {\bibfnamefont {P.}~\bibnamefont {Jarillo-Herrero}},\ }\href {\doibase
  10.1038/s41586-021-03192-0} {\bibfield  {journal} {\bibinfo  {journal}
  {Nature}\ }\textbf {\bibinfo {volume} {590}},\ \bibinfo {pages} {249}
  (\bibinfo {year} {2021}{\natexlab{a}})}\BibitemShut {NoStop}%
\bibitem [{\citenamefont {Hao}\ \emph {et~al.}(2021)\citenamefont {Hao},
  \citenamefont {Zimmerman}, \citenamefont {Ledwith}, \citenamefont {Khalaf},
  \citenamefont {Najafabadi}, \citenamefont {Watanabe}, \citenamefont
  {Taniguchi}, \citenamefont {Vishwanath},\ and\ \citenamefont
  {Kim}}]{hao2020electric}%
  \BibitemOpen
  \bibfield  {author} {\bibinfo {author} {\bibfnamefont {Z.}~\bibnamefont
  {Hao}}, \bibinfo {author} {\bibfnamefont {A.~M.}\ \bibnamefont {Zimmerman}},
  \bibinfo {author} {\bibfnamefont {P.}~\bibnamefont {Ledwith}}, \bibinfo
  {author} {\bibfnamefont {E.}~\bibnamefont {Khalaf}}, \bibinfo {author}
  {\bibfnamefont {D.~H.}\ \bibnamefont {Najafabadi}}, \bibinfo {author}
  {\bibfnamefont {K.}~\bibnamefont {Watanabe}}, \bibinfo {author}
  {\bibfnamefont {T.}~\bibnamefont {Taniguchi}}, \bibinfo {author}
  {\bibfnamefont {A.}~\bibnamefont {Vishwanath}}, \ and\ \bibinfo {author}
  {\bibfnamefont {P.}~\bibnamefont {Kim}},\ }\href {\doibase
  10.1126/science.abg0399} {\bibfield  {journal} {\bibinfo  {journal}
  {Science}\ }\textbf {\bibinfo {volume} {371}},\ \bibinfo {pages} {1133}
  (\bibinfo {year} {2021})}\BibitemShut {NoStop}%
\bibitem [{\citenamefont {Lee}\ \emph {et~al.}(2019)\citenamefont {Lee},
  \citenamefont {Khalaf}, \citenamefont {Liu}, \citenamefont {Liu},
  \citenamefont {Hao}, \citenamefont {Kim},\ and\ \citenamefont
  {Vishwanath}}]{Lee2019}%
  \BibitemOpen
  \bibfield  {author} {\bibinfo {author} {\bibfnamefont {J.~Y.}\ \bibnamefont
  {Lee}}, \bibinfo {author} {\bibfnamefont {E.}~\bibnamefont {Khalaf}},
  \bibinfo {author} {\bibfnamefont {S.}~\bibnamefont {Liu}}, \bibinfo {author}
  {\bibfnamefont {X.}~\bibnamefont {Liu}}, \bibinfo {author} {\bibfnamefont
  {Z.}~\bibnamefont {Hao}}, \bibinfo {author} {\bibfnamefont {P.}~\bibnamefont
  {Kim}}, \ and\ \bibinfo {author} {\bibfnamefont {A.}~\bibnamefont
  {Vishwanath}},\ }\href {\doibase 10.1038/s41467-019-12981-1} {\bibfield
  {journal} {\bibinfo  {journal} {Nat. Commun.}\ }\textbf {\bibinfo {volume}
  {10}},\ \bibinfo {pages} {5333} (\bibinfo {year} {2019})}\BibitemShut
  {NoStop}%
\bibitem [{\citenamefont {Chebrolu}\ \emph {et~al.}(2019)\citenamefont
  {Chebrolu}, \citenamefont {Chittari},\ and\ \citenamefont
  {Jung}}]{Chebrolu2019}%
  \BibitemOpen
  \bibfield  {author} {\bibinfo {author} {\bibfnamefont {N.~R.}\ \bibnamefont
  {Chebrolu}}, \bibinfo {author} {\bibfnamefont {B.~L.}\ \bibnamefont
  {Chittari}}, \ and\ \bibinfo {author} {\bibfnamefont {J.}~\bibnamefont
  {Jung}},\ }\href {\doibase 10.1103/PhysRevB.99.235417} {\bibfield  {journal}
  {\bibinfo  {journal} {Phys. Rev. B}\ }\textbf {\bibinfo {volume} {99}},\
  \bibinfo {pages} {235417} (\bibinfo {year} {2019})}\BibitemShut {NoStop}%
\bibitem [{\citenamefont {Choi}\ and\ \citenamefont {Choi}(2019)}]{Choi2019}%
  \BibitemOpen
  \bibfield  {author} {\bibinfo {author} {\bibfnamefont {Y.~W.}\ \bibnamefont
  {Choi}}\ and\ \bibinfo {author} {\bibfnamefont {H.~J.}\ \bibnamefont
  {Choi}},\ }\href {\doibase 10.1103/PhysRevB.100.201402} {\bibfield  {journal}
  {\bibinfo  {journal} {Phys. Rev. B}\ }\textbf {\bibinfo {volume} {100}},\
  \bibinfo {pages} {201402(R)} (\bibinfo {year} {2019})}\BibitemShut {NoStop}%
\bibitem [{\citenamefont {Koshino}(2019)}]{Koshino2019}%
  \BibitemOpen
  \bibfield  {author} {\bibinfo {author} {\bibfnamefont {M.}~\bibnamefont
  {Koshino}},\ }\href {\doibase 10.1103/PhysRevB.99.235406} {\bibfield
  {journal} {\bibinfo  {journal} {Phys. Rev. B}\ }\textbf {\bibinfo {volume}
  {99}},\ \bibinfo {pages} {235406} (\bibinfo {year} {2019})}\BibitemShut
  {NoStop}%
\bibitem [{\citenamefont {Burg}\ \emph {et~al.}(2019)\citenamefont {Burg},
  \citenamefont {Zhu}, \citenamefont {Taniguchi}, \citenamefont {Watanabe},
  \citenamefont {MacDonald},\ and\ \citenamefont {Tutuc}}]{Burg2019}%
  \BibitemOpen
  \bibfield  {author} {\bibinfo {author} {\bibfnamefont {G.~W.}\ \bibnamefont
  {Burg}}, \bibinfo {author} {\bibfnamefont {J.}~\bibnamefont {Zhu}}, \bibinfo
  {author} {\bibfnamefont {T.}~\bibnamefont {Taniguchi}}, \bibinfo {author}
  {\bibfnamefont {K.}~\bibnamefont {Watanabe}}, \bibinfo {author}
  {\bibfnamefont {A.~H.}\ \bibnamefont {MacDonald}}, \ and\ \bibinfo {author}
  {\bibfnamefont {E.}~\bibnamefont {Tutuc}},\ }\href {\doibase
  10.1103/PhysRevLett.123.197702} {\bibfield  {journal} {\bibinfo  {journal}
  {Phys. Rev. Lett.}\ }\textbf {\bibinfo {volume} {123}},\ \bibinfo {pages}
  {197702} (\bibinfo {year} {2019})}\BibitemShut {NoStop}%
\bibitem [{\citenamefont {Cao}\ \emph {et~al.}(2020{\natexlab{b}})\citenamefont
  {Cao}, \citenamefont {Rodan-Legrain}, \citenamefont {Rubies-Bigorda},
  \citenamefont {Park}, \citenamefont {Watanabe}, \citenamefont {Taniguchi},\
  and\ \citenamefont {Jarillo-Herrero}}]{Cao2019a}%
  \BibitemOpen
  \bibfield  {author} {\bibinfo {author} {\bibfnamefont {Y.}~\bibnamefont
  {Cao}}, \bibinfo {author} {\bibfnamefont {D.}~\bibnamefont {Rodan-Legrain}},
  \bibinfo {author} {\bibfnamefont {O.}~\bibnamefont {Rubies-Bigorda}},
  \bibinfo {author} {\bibfnamefont {J.~M.}\ \bibnamefont {Park}}, \bibinfo
  {author} {\bibfnamefont {K.}~\bibnamefont {Watanabe}}, \bibinfo {author}
  {\bibfnamefont {T.}~\bibnamefont {Taniguchi}}, \ and\ \bibinfo {author}
  {\bibfnamefont {P.}~\bibnamefont {Jarillo-Herrero}},\ }\href
  {https://www.nature.com/articles/s41586-020-2260-6} {\bibfield  {journal}
  {\bibinfo  {journal} {Nature}\ }\textbf {\bibinfo {volume} {583}},\ \bibinfo
  {pages} {215} (\bibinfo {year} {2020}{\natexlab{b}})}\BibitemShut {NoStop}%
\bibitem [{\citenamefont {Shen}\ \emph {et~al.}(2020)\citenamefont {Shen},
  \citenamefont {Li}, \citenamefont {Wang}, \citenamefont {Zhao}, \citenamefont
  {Tang}, \citenamefont {Liu}, \citenamefont {Tian}, \citenamefont {Chu},
  \citenamefont {Watanabe}, \citenamefont {Taniguchi}, \citenamefont {Yang},
  \citenamefont {Meng}, \citenamefont {Shi},\ and\ \citenamefont
  {Zhang}}]{Shen2019}%
  \BibitemOpen
  \bibfield  {author} {\bibinfo {author} {\bibfnamefont {C.}~\bibnamefont
  {Shen}}, \bibinfo {author} {\bibfnamefont {N.}~\bibnamefont {Li}}, \bibinfo
  {author} {\bibfnamefont {S.}~\bibnamefont {Wang}}, \bibinfo {author}
  {\bibfnamefont {Y.}~\bibnamefont {Zhao}}, \bibinfo {author} {\bibfnamefont
  {J.}~\bibnamefont {Tang}}, \bibinfo {author} {\bibfnamefont {J.}~\bibnamefont
  {Liu}}, \bibinfo {author} {\bibfnamefont {J.}~\bibnamefont {Tian}}, \bibinfo
  {author} {\bibfnamefont {Y.}~\bibnamefont {Chu}}, \bibinfo {author}
  {\bibfnamefont {K.}~\bibnamefont {Watanabe}}, \bibinfo {author}
  {\bibfnamefont {T.}~\bibnamefont {Taniguchi}}, \bibinfo {author}
  {\bibfnamefont {R.}~\bibnamefont {Yang}}, \bibinfo {author} {\bibfnamefont
  {Z.~Y.}\ \bibnamefont {Meng}}, \bibinfo {author} {\bibfnamefont
  {D.}~\bibnamefont {Shi}}, \ and\ \bibinfo {author} {\bibfnamefont
  {G.}~\bibnamefont {Zhang}},\ }\href
  {https://www.nature.com/articles/s41567-020-0825-9} {\bibfield  {journal}
  {\bibinfo  {journal} {Nat. Phys.}\ }\textbf {\bibinfo {volume} {16}},\
  \bibinfo {pages} {520} (\bibinfo {year} {2020})}\BibitemShut {NoStop}%
\bibitem [{\citenamefont {Liu}\ \emph {et~al.}(2020)\citenamefont {Liu},
  \citenamefont {Hao}, \citenamefont {Khalaf}, \citenamefont {Lee},
  \citenamefont {Ronen}, \citenamefont {Yoo}, \citenamefont {Najafabadi},
  \citenamefont {Watanabe}, \citenamefont {Taniguchi}, \citenamefont
  {Vishwanath},\ and\ \citenamefont {Kim}}]{Liu2019}%
  \BibitemOpen
  \bibfield  {author} {\bibinfo {author} {\bibfnamefont {X.}~\bibnamefont
  {Liu}}, \bibinfo {author} {\bibfnamefont {Z.}~\bibnamefont {Hao}}, \bibinfo
  {author} {\bibfnamefont {E.}~\bibnamefont {Khalaf}}, \bibinfo {author}
  {\bibfnamefont {J.~Y.}\ \bibnamefont {Lee}}, \bibinfo {author} {\bibfnamefont
  {Y.}~\bibnamefont {Ronen}}, \bibinfo {author} {\bibfnamefont
  {H.}~\bibnamefont {Yoo}}, \bibinfo {author} {\bibfnamefont {D.~H.}\
  \bibnamefont {Najafabadi}}, \bibinfo {author} {\bibfnamefont
  {K.}~\bibnamefont {Watanabe}}, \bibinfo {author} {\bibfnamefont
  {T.}~\bibnamefont {Taniguchi}}, \bibinfo {author} {\bibfnamefont
  {A.}~\bibnamefont {Vishwanath}}, \ and\ \bibinfo {author} {\bibfnamefont
  {P.}~\bibnamefont {Kim}},\ }\href
  {https://www.nature.com/articles/s41586-020-2458-7} {\bibfield  {journal}
  {\bibinfo  {journal} {Nature}\ }\textbf {\bibinfo {volume} {583}},\ \bibinfo
  {pages} {221} (\bibinfo {year} {2020})}\BibitemShut {NoStop}%
\bibitem [{\citenamefont {Chittari}\ \emph {et~al.}(2019)\citenamefont
  {Chittari}, \citenamefont {Chen}, \citenamefont {Zhang}, \citenamefont
  {Wang},\ and\ \citenamefont {Jung}}]{Chittari2019}%
  \BibitemOpen
  \bibfield  {author} {\bibinfo {author} {\bibfnamefont {B.~L.}\ \bibnamefont
  {Chittari}}, \bibinfo {author} {\bibfnamefont {G.}~\bibnamefont {Chen}},
  \bibinfo {author} {\bibfnamefont {Y.}~\bibnamefont {Zhang}}, \bibinfo
  {author} {\bibfnamefont {F.}~\bibnamefont {Wang}}, \ and\ \bibinfo {author}
  {\bibfnamefont {J.}~\bibnamefont {Jung}},\ }\href {\doibase
  10.1103/PhysRevLett.122.016401} {\bibfield  {journal} {\bibinfo  {journal}
  {Phys. Rev. Lett.}\ }\textbf {\bibinfo {volume} {122}},\ \bibinfo {pages}
  {016401} (\bibinfo {year} {2019})}\BibitemShut {NoStop}%
\bibitem [{\citenamefont {Su\'arez~Morell}\ \emph {et~al.}(2013)\citenamefont
  {Su\'arez~Morell}, \citenamefont {Pacheco}, \citenamefont {Chico},\ and\
  \citenamefont {Brey}}]{SuarezMorell2013}%
  \BibitemOpen
  \bibfield  {author} {\bibinfo {author} {\bibfnamefont {E.}~\bibnamefont
  {Su\'arez~Morell}}, \bibinfo {author} {\bibfnamefont {M.}~\bibnamefont
  {Pacheco}}, \bibinfo {author} {\bibfnamefont {L.}~\bibnamefont {Chico}}, \
  and\ \bibinfo {author} {\bibfnamefont {L.}~\bibnamefont {Brey}},\ }\href
  {\doibase 10.1103/PhysRevB.87.125414} {\bibfield  {journal} {\bibinfo
  {journal} {Phys. Rev. B}\ }\textbf {\bibinfo {volume} {87}},\ \bibinfo
  {pages} {125414} (\bibinfo {year} {2013})}\BibitemShut {NoStop}%
\bibitem [{\citenamefont {Li}\ \emph {et~al.}(2019)\citenamefont {Li},
  \citenamefont {Wu},\ and\ \citenamefont {MacDonald}}]{Li2019}%
  \BibitemOpen
  \bibfield  {author} {\bibinfo {author} {\bibfnamefont {X.}~\bibnamefont
  {Li}}, \bibinfo {author} {\bibfnamefont {F.}~\bibnamefont {Wu}}, \ and\
  \bibinfo {author} {\bibfnamefont {A.~H.}\ \bibnamefont {MacDonald}},\ }\href
  {http://arxiv.org/abs/1907.12338} {\  (\bibinfo {year} {2019})},\ \Eprint
  {http://arxiv.org/abs/1907.12338} {arXiv:1907.12338} \BibitemShut {NoStop}%
\bibitem [{\citenamefont {Szendrő}\ \emph {et~al.}(2020)\citenamefont
  {Szendrő}, \citenamefont {Süle}, \citenamefont {Dobrik},\ and\
  \citenamefont {Tapasztó}}]{szendr2020ultraflat}%
  \BibitemOpen
  \bibfield  {author} {\bibinfo {author} {\bibfnamefont {M.}~\bibnamefont
  {Szendrő}}, \bibinfo {author} {\bibfnamefont {P.}~\bibnamefont {Süle}},
  \bibinfo {author} {\bibfnamefont {G.}~\bibnamefont {Dobrik}}, \ and\ \bibinfo
  {author} {\bibfnamefont {L.}~\bibnamefont {Tapasztó}},\ }\href@noop {} {\
  (\bibinfo {year} {2020})},\ \Eprint {http://arxiv.org/abs/2001.11462}
  {arXiv:2001.11462} \BibitemShut {NoStop}%
\bibitem [{\citenamefont {Carr}\ \emph {et~al.}(2020)\citenamefont {Carr},
  \citenamefont {Li}, \citenamefont {Zhu}, \citenamefont {Kaxiras},
  \citenamefont {Sachdev},\ and\ \citenamefont {Kruchkov}}]{Carr2019a}%
  \BibitemOpen
  \bibfield  {author} {\bibinfo {author} {\bibfnamefont {S.}~\bibnamefont
  {Carr}}, \bibinfo {author} {\bibfnamefont {C.}~\bibnamefont {Li}}, \bibinfo
  {author} {\bibfnamefont {Z.}~\bibnamefont {Zhu}}, \bibinfo {author}
  {\bibfnamefont {E.}~\bibnamefont {Kaxiras}}, \bibinfo {author} {\bibfnamefont
  {S.}~\bibnamefont {Sachdev}}, \ and\ \bibinfo {author} {\bibfnamefont
  {A.}~\bibnamefont {Kruchkov}},\ }\href {\doibase
  https://doi.org/10.1021/acs.nanolett.9b04979} {\bibfield  {journal} {\bibinfo
   {journal} {Nano Lett.}\ }\textbf {\bibinfo {volume} {20}},\ \bibinfo {pages}
  {3030} (\bibinfo {year} {2020})}\BibitemShut {NoStop}%
\bibitem [{\citenamefont {Ma}\ \emph {et~al.}(2021)\citenamefont {Ma},
  \citenamefont {Li}, \citenamefont {Zheng}, \citenamefont {Xiao},
  \citenamefont {Jiang}, \citenamefont {Gao},\ and\ \citenamefont
  {Xie}}]{Ma2019}%
  \BibitemOpen
  \bibfield  {author} {\bibinfo {author} {\bibfnamefont {Z.}~\bibnamefont
  {Ma}}, \bibinfo {author} {\bibfnamefont {S.}~\bibnamefont {Li}}, \bibinfo
  {author} {\bibfnamefont {Y.-W.}\ \bibnamefont {Zheng}}, \bibinfo {author}
  {\bibfnamefont {M.-M.}\ \bibnamefont {Xiao}}, \bibinfo {author}
  {\bibfnamefont {H.}~\bibnamefont {Jiang}}, \bibinfo {author} {\bibfnamefont
  {J.-H.}\ \bibnamefont {Gao}}, \ and\ \bibinfo {author} {\bibfnamefont
  {X.}~\bibnamefont {Xie}},\ }\href {\doibase
  https://doi.org/10.1016/j.scib.2020.10.004} {\bibfield  {journal} {\bibinfo
  {journal} {Science Bulletin}\ }\textbf {\bibinfo {volume} {66}},\ \bibinfo
  {pages} {18} (\bibinfo {year} {2021})}\BibitemShut {NoStop}%
\bibitem [{\citenamefont {Xu}\ \emph {et~al.}(2021)\citenamefont {Xu},
  \citenamefont {Ezzi}, \citenamefont {Balakrishnan}, \citenamefont
  {Garcia-Ruiz}, \citenamefont {Tsim}, \citenamefont {Mullan}, \citenamefont
  {Barrier}, \citenamefont {Xin}, \citenamefont {Piot}, \citenamefont
  {Taniguchi}, \citenamefont {Watanabe}, \citenamefont {Carvalho},
  \citenamefont {Mishchenko}, \citenamefont {Geim}, \citenamefont {Fal’ko},
  \citenamefont {Adam}, \citenamefont {Neto}, \citenamefont {Novoselov},\ and\
  \citenamefont {Shi}}]{tmbg1}%
  \BibitemOpen
  \bibfield  {author} {\bibinfo {author} {\bibfnamefont {S.}~\bibnamefont
  {Xu}}, \bibinfo {author} {\bibfnamefont {M.~M.~A.}\ \bibnamefont {Ezzi}},
  \bibinfo {author} {\bibfnamefont {N.}~\bibnamefont {Balakrishnan}}, \bibinfo
  {author} {\bibfnamefont {A.}~\bibnamefont {Garcia-Ruiz}}, \bibinfo {author}
  {\bibfnamefont {B.}~\bibnamefont {Tsim}}, \bibinfo {author} {\bibfnamefont
  {C.}~\bibnamefont {Mullan}}, \bibinfo {author} {\bibfnamefont
  {J.}~\bibnamefont {Barrier}}, \bibinfo {author} {\bibfnamefont
  {N.}~\bibnamefont {Xin}}, \bibinfo {author} {\bibfnamefont {B.~A.}\
  \bibnamefont {Piot}}, \bibinfo {author} {\bibfnamefont {T.}~\bibnamefont
  {Taniguchi}}, \bibinfo {author} {\bibfnamefont {K.}~\bibnamefont {Watanabe}},
  \bibinfo {author} {\bibfnamefont {A.}~\bibnamefont {Carvalho}}, \bibinfo
  {author} {\bibfnamefont {A.}~\bibnamefont {Mishchenko}}, \bibinfo {author}
  {\bibfnamefont {A.~K.}\ \bibnamefont {Geim}}, \bibinfo {author}
  {\bibfnamefont {V.~I.}\ \bibnamefont {Fal’ko}}, \bibinfo {author}
  {\bibfnamefont {S.}~\bibnamefont {Adam}}, \bibinfo {author} {\bibfnamefont
  {A.~H.~C.}\ \bibnamefont {Neto}}, \bibinfo {author} {\bibfnamefont {K.~S.}\
  \bibnamefont {Novoselov}}, \ and\ \bibinfo {author} {\bibfnamefont
  {Y.}~\bibnamefont {Shi}},\ }\href@noop {} {\bibfield  {journal} {\bibinfo
  {journal} {Nat. Phys.}\ }\textbf {\bibinfo {volume} {17}},\ \bibinfo {pages}
  {619} (\bibinfo {year} {2021})}\BibitemShut {NoStop}%
\bibitem [{\citenamefont {Chen}\ \emph {et~al.}(2021)\citenamefont {Chen},
  \citenamefont {He}, \citenamefont {Zhang}, \citenamefont {Hsieh},
  \citenamefont {Fei}, \citenamefont {Watanabe}, \citenamefont {Taniguchi},
  \citenamefont {Cobden}, \citenamefont {Xu}, \citenamefont {Dean},\ and\
  \citenamefont {Yankowitz}}]{tmbg2}%
  \BibitemOpen
  \bibfield  {author} {\bibinfo {author} {\bibfnamefont {S.}~\bibnamefont
  {Chen}}, \bibinfo {author} {\bibfnamefont {M.}~\bibnamefont {He}}, \bibinfo
  {author} {\bibfnamefont {Y.-H.}\ \bibnamefont {Zhang}}, \bibinfo {author}
  {\bibfnamefont {V.}~\bibnamefont {Hsieh}}, \bibinfo {author} {\bibfnamefont
  {Z.}~\bibnamefont {Fei}}, \bibinfo {author} {\bibfnamefont {K.}~\bibnamefont
  {Watanabe}}, \bibinfo {author} {\bibfnamefont {T.}~\bibnamefont {Taniguchi}},
  \bibinfo {author} {\bibfnamefont {D.~H.}\ \bibnamefont {Cobden}}, \bibinfo
  {author} {\bibfnamefont {X.}~\bibnamefont {Xu}}, \bibinfo {author}
  {\bibfnamefont {C.~R.}\ \bibnamefont {Dean}}, \ and\ \bibinfo {author}
  {\bibfnamefont {M.}~\bibnamefont {Yankowitz}},\ }\href@noop {} {\bibfield
  {journal} {\bibinfo  {journal} {Nat. Phys.}\ }\textbf {\bibinfo {volume}
  {17}},\ \bibinfo {pages} {374} (\bibinfo {year} {2021})}\BibitemShut
  {NoStop}%
\bibitem [{\citenamefont {Polshyn}\ \emph {et~al.}(2020)\citenamefont
  {Polshyn}, \citenamefont {Zhu}, \citenamefont {Kumar}, \citenamefont {Zhang},
  \citenamefont {Yang}, \citenamefont {Tschirhart}, \citenamefont {Serlin},
  \citenamefont {Watanabe}, \citenamefont {Taniguchi}, \citenamefont
  {MacDonald},\ and\ \citenamefont {Young}}]{tmbg3}%
  \BibitemOpen
  \bibfield  {author} {\bibinfo {author} {\bibfnamefont {H.}~\bibnamefont
  {Polshyn}}, \bibinfo {author} {\bibfnamefont {J.}~\bibnamefont {Zhu}},
  \bibinfo {author} {\bibfnamefont {M.~A.}\ \bibnamefont {Kumar}}, \bibinfo
  {author} {\bibfnamefont {Y.}~\bibnamefont {Zhang}}, \bibinfo {author}
  {\bibfnamefont {F.}~\bibnamefont {Yang}}, \bibinfo {author} {\bibfnamefont
  {C.~L.}\ \bibnamefont {Tschirhart}}, \bibinfo {author} {\bibfnamefont
  {M.}~\bibnamefont {Serlin}}, \bibinfo {author} {\bibfnamefont
  {K.}~\bibnamefont {Watanabe}}, \bibinfo {author} {\bibfnamefont
  {T.}~\bibnamefont {Taniguchi}}, \bibinfo {author} {\bibfnamefont {A.~H.}\
  \bibnamefont {MacDonald}}, \ and\ \bibinfo {author} {\bibfnamefont {A.~F.}\
  \bibnamefont {Young}},\ }\href@noop {} {\bibfield  {journal} {\bibinfo
  {journal} {Nature}\ }\textbf {\bibinfo {volume} {588}},\ \bibinfo {pages}
  {66} (\bibinfo {year} {2020})}\BibitemShut {NoStop}%
\bibitem [{\citenamefont {Park}\ \emph {et~al.}(2020)\citenamefont {Park},
  \citenamefont {Chittari},\ and\ \citenamefont {Jung}}]{Park2020}%
  \BibitemOpen
  \bibfield  {author} {\bibinfo {author} {\bibfnamefont {Y.}~\bibnamefont
  {Park}}, \bibinfo {author} {\bibfnamefont {B.~L.}\ \bibnamefont {Chittari}},
  \ and\ \bibinfo {author} {\bibfnamefont {J.}~\bibnamefont {Jung}},\ }\href
  {\doibase 10.1103/PhysRevB.102.035411} {\bibfield  {journal} {\bibinfo
  {journal} {Phys. Rev. B}\ }\textbf {\bibinfo {volume} {102}},\ \bibinfo
  {pages} {035411} (\bibinfo {year} {2020})}\BibitemShut {NoStop}%
\bibitem [{\citenamefont {Lei}\ \emph {et~al.}(2020)\citenamefont {Lei},
  \citenamefont {Linhart}, \citenamefont {Qin}, \citenamefont {Libisch},\ and\
  \citenamefont {MacDonald}}]{lei2020mirror}%
  \BibitemOpen
  \bibfield  {author} {\bibinfo {author} {\bibfnamefont {C.}~\bibnamefont
  {Lei}}, \bibinfo {author} {\bibfnamefont {L.}~\bibnamefont {Linhart}},
  \bibinfo {author} {\bibfnamefont {W.}~\bibnamefont {Qin}}, \bibinfo {author}
  {\bibfnamefont {F.}~\bibnamefont {Libisch}}, \ and\ \bibinfo {author}
  {\bibfnamefont {A.~H.}\ \bibnamefont {MacDonald}},\ }\href@noop {} {\
  (\bibinfo {year} {2020})},\ \Eprint {http://arxiv.org/abs/2010.05787}
  {arXiv:2010.05787} \BibitemShut {NoStop}%
\bibitem [{\citenamefont {Wu}\ \emph {et~al.}(2021)\citenamefont {Wu},
  \citenamefont {Zhan},\ and\ \citenamefont {Yuan}}]{wu2020lattice}%
  \BibitemOpen
  \bibfield  {author} {\bibinfo {author} {\bibfnamefont {Z.}~\bibnamefont
  {Wu}}, \bibinfo {author} {\bibfnamefont {Z.}~\bibnamefont {Zhan}}, \ and\
  \bibinfo {author} {\bibfnamefont {S.}~\bibnamefont {Yuan}},\ }\href
  {https://doi.org/10.1007/s11433-020-1690-4} {\bibfield  {journal} {\bibinfo
  {journal} {Sci. China Phys. Mech. Astron.}\ }\textbf {\bibinfo {volume}
  {64}},\ \bibinfo {pages} {267811} (\bibinfo {year} {2021})}\BibitemShut
  {NoStop}%
\bibitem [{\citenamefont {Mora}\ \emph {et~al.}(2019)\citenamefont {Mora},
  \citenamefont {Regnault},\ and\ \citenamefont {Bernevig}}]{Mora2019}%
  \BibitemOpen
  \bibfield  {author} {\bibinfo {author} {\bibfnamefont {C.}~\bibnamefont
  {Mora}}, \bibinfo {author} {\bibfnamefont {N.}~\bibnamefont {Regnault}}, \
  and\ \bibinfo {author} {\bibfnamefont {B.~A.}\ \bibnamefont {Bernevig}},\
  }\href {\doibase 10.1103/PhysRevLett.123.026402} {\bibfield  {journal}
  {\bibinfo  {journal} {Phys. Rev. Lett.}\ }\textbf {\bibinfo {volume} {123}},\
  \bibinfo {pages} {026402} (\bibinfo {year} {2019})}\BibitemShut {NoStop}%
\bibitem [{\citenamefont {Zuo}\ \emph {et~al.}(2018)\citenamefont {Zuo},
  \citenamefont {Qiao}, \citenamefont {Ma}, \citenamefont {Yin}, \citenamefont
  {Sun}, \citenamefont {Zhang}, \citenamefont {Guan},\ and\ \citenamefont
  {He}}]{Zuo2018}%
  \BibitemOpen
  \bibfield  {author} {\bibinfo {author} {\bibfnamefont {W.-J.}\ \bibnamefont
  {Zuo}}, \bibinfo {author} {\bibfnamefont {J.-B.}\ \bibnamefont {Qiao}},
  \bibinfo {author} {\bibfnamefont {D.-L.}\ \bibnamefont {Ma}}, \bibinfo
  {author} {\bibfnamefont {L.-J.}\ \bibnamefont {Yin}}, \bibinfo {author}
  {\bibfnamefont {G.}~\bibnamefont {Sun}}, \bibinfo {author} {\bibfnamefont
  {J.-Y.}\ \bibnamefont {Zhang}}, \bibinfo {author} {\bibfnamefont {L.-Y.}\
  \bibnamefont {Guan}}, \ and\ \bibinfo {author} {\bibfnamefont
  {L.}~\bibnamefont {He}},\ }\href {\doibase 10.1103/PhysRevB.97.035440}
  {\bibfield  {journal} {\bibinfo  {journal} {Phys. Rev. B}\ }\textbf {\bibinfo
  {volume} {97}},\ \bibinfo {pages} {035440} (\bibinfo {year}
  {2018})}\BibitemShut {NoStop}%
\bibitem [{\citenamefont {Khalaf}\ \emph {et~al.}(2019)\citenamefont {Khalaf},
  \citenamefont {Kruchkov}, \citenamefont {Tarnopolsky},\ and\ \citenamefont
  {Vishwanath}}]{Khalaf2019}%
  \BibitemOpen
  \bibfield  {author} {\bibinfo {author} {\bibfnamefont {E.}~\bibnamefont
  {Khalaf}}, \bibinfo {author} {\bibfnamefont {A.~J.}\ \bibnamefont
  {Kruchkov}}, \bibinfo {author} {\bibfnamefont {G.}~\bibnamefont
  {Tarnopolsky}}, \ and\ \bibinfo {author} {\bibfnamefont {A.}~\bibnamefont
  {Vishwanath}},\ }\href {\doibase 10.1103/PhysRevB.100.085109} {\bibfield
  {journal} {\bibinfo  {journal} {Phys. Rev. B}\ }\textbf {\bibinfo {volume}
  {100}},\ \bibinfo {pages} {085109} (\bibinfo {year} {2019})}\BibitemShut
  {NoStop}%
\bibitem [{\citenamefont {Lopez-Bezanilla}\ and\ \citenamefont
  {Lado}(2020)}]{tb_tTG}%
  \BibitemOpen
  \bibfield  {author} {\bibinfo {author} {\bibfnamefont {A.}~\bibnamefont
  {Lopez-Bezanilla}}\ and\ \bibinfo {author} {\bibfnamefont {J.~L.}\
  \bibnamefont {Lado}},\ }\href {\doibase 10.1103/PhysRevResearch.2.033357}
  {\bibfield  {journal} {\bibinfo  {journal} {Phys. Rev. Research}\ }\textbf
  {\bibinfo {volume} {2}},\ \bibinfo {pages} {033357} (\bibinfo {year}
  {2020})}\BibitemShut {NoStop}%
\bibitem [{\citenamefont {C\ifmmode \u{a}\else \u{a}\fi{}lug\ifmmode~\u{a}\else
  \u{a}\fi{}ru}\ \emph {et~al.}(2021)\citenamefont {C\ifmmode \u{a}\else
  \u{a}\fi{}lug\ifmmode~\u{a}\else \u{a}\fi{}ru}, \citenamefont {Xie},
  \citenamefont {Song}, \citenamefont {Lian}, \citenamefont {Regnault},\ and\
  \citenamefont {Bernevig}}]{Dumitru2021}%
  \BibitemOpen
  \bibfield  {author} {\bibinfo {author} {\bibfnamefont {D.}~\bibnamefont
  {C\ifmmode \u{a}\else \u{a}\fi{}lug\ifmmode~\u{a}\else \u{a}\fi{}ru}},
  \bibinfo {author} {\bibfnamefont {F.}~\bibnamefont {Xie}}, \bibinfo {author}
  {\bibfnamefont {Z.-D.}\ \bibnamefont {Song}}, \bibinfo {author}
  {\bibfnamefont {B.}~\bibnamefont {Lian}}, \bibinfo {author} {\bibfnamefont
  {N.}~\bibnamefont {Regnault}}, \ and\ \bibinfo {author} {\bibfnamefont
  {B.~A.}\ \bibnamefont {Bernevig}},\ }\href {\doibase
  10.1103/PhysRevB.103.195411} {\bibfield  {journal} {\bibinfo  {journal}
  {Phys. Rev. B}\ }\textbf {\bibinfo {volume} {103}},\ \bibinfo {pages}
  {195411} (\bibinfo {year} {2021})}\BibitemShut {NoStop}%
\bibitem [{\citenamefont {Shin}\ \emph {et~al.}(2021)\citenamefont {Shin},
  \citenamefont {Chittari},\ and\ \citenamefont {Jung}}]{Shin2021}%
  \BibitemOpen
  \bibfield  {author} {\bibinfo {author} {\bibfnamefont {J.}~\bibnamefont
  {Shin}}, \bibinfo {author} {\bibfnamefont {B.~L.}\ \bibnamefont {Chittari}},
  \ and\ \bibinfo {author} {\bibfnamefont {J.}~\bibnamefont {Jung}},\ }\href
  {\doibase 10.1103/PhysRevB.104.045413} {\bibfield  {journal} {\bibinfo
  {journal} {Phys. Rev. B}\ }\textbf {\bibinfo {volume} {104}},\ \bibinfo
  {pages} {045413} (\bibinfo {year} {2021})}\BibitemShut {NoStop}%
\bibitem [{\citenamefont {Min}\ \emph {et~al.}(2007)\citenamefont {Min},
  \citenamefont {Sahu}, \citenamefont {Banerjee},\ and\ \citenamefont
  {MacDonald}}]{min2007}%
  \BibitemOpen
  \bibfield  {author} {\bibinfo {author} {\bibfnamefont {H.}~\bibnamefont
  {Min}}, \bibinfo {author} {\bibfnamefont {B.}~\bibnamefont {Sahu}}, \bibinfo
  {author} {\bibfnamefont {S.~K.}\ \bibnamefont {Banerjee}}, \ and\ \bibinfo
  {author} {\bibfnamefont {A.~H.}~\bibnamefont {MacDonald}},\ }\href@noop {}
  {\bibfield  {journal} {\bibinfo  {journal} {Physical Review B}\ }\textbf
  {\bibinfo {volume} {75}},\ \bibinfo {pages} {155115} (\bibinfo {year}
  {2007})}\BibitemShut {NoStop}%
\bibitem [{\citenamefont {Bistritzer}\ and\ \citenamefont
  {MacDonald}(2011)}]{Bistritzer2010b}%
  \BibitemOpen
  \bibfield  {author} {\bibinfo {author} {\bibfnamefont {R.}~\bibnamefont
  {Bistritzer}}\ and\ \bibinfo {author} {\bibfnamefont {A.~H.}\ \bibnamefont
  {MacDonald}},\ }\href {\doibase 10.1073/pnas.1108174108} {\bibfield
  {journal} {\bibinfo  {journal} {PNAS}\ }\textbf {\bibinfo {volume} {108}},\
  \bibinfo {pages} {12233} (\bibinfo {year} {2011})}\BibitemShut {NoStop}%
\bibitem [{\citenamefont {Jung}\ \emph {et~al.}(2014)\citenamefont {Jung},
  \citenamefont {Raoux}, \citenamefont {Qiao},\ and\ \citenamefont
  {MacDonald}}]{Jung2014}%
  \BibitemOpen
  \bibfield  {author} {\bibinfo {author} {\bibfnamefont {J.}~\bibnamefont
  {Jung}}, \bibinfo {author} {\bibfnamefont {A.}~\bibnamefont {Raoux}},
  \bibinfo {author} {\bibfnamefont {Z.}~\bibnamefont {Qiao}}, \ and\ \bibinfo
  {author} {\bibfnamefont {A.~H.}\ \bibnamefont {MacDonald}},\ }\href {\doibase
  10.1103/PhysRevB.89.205414} {\bibfield  {journal} {\bibinfo  {journal} {Phys.
  Rev. B}\ }\textbf {\bibinfo {volume} {89}},\ \bibinfo {pages} {205414}
  (\bibinfo {year} {2014})}\BibitemShut {NoStop}%
\bibitem [{\citenamefont {Jung}\ and\ \citenamefont
  {MacDonald}(2014)}]{Jung2014a}%
  \BibitemOpen
  \bibfield  {author} {\bibinfo {author} {\bibfnamefont {J.}~\bibnamefont
  {Jung}}\ and\ \bibinfo {author} {\bibfnamefont {A.~H.}\ \bibnamefont
  {MacDonald}},\ }\href {\doibase 10.1103/PhysRevB.89.035405} {\bibfield
  {journal} {\bibinfo  {journal} {Phys. Rev. B}\ }\textbf {\bibinfo {volume}
  {89}},\ \bibinfo {pages} {035405} (\bibinfo {year} {2014})}\BibitemShut
  {NoStop}%
\bibitem [{\citenamefont {Park}\ \emph
  {et~al.}(2021{\natexlab{b}})\citenamefont {Park}, \citenamefont {Cao},
  \citenamefont {Xia}, \citenamefont {Sun}, \citenamefont {Watanabe},
  \citenamefont {Taniguchi},\ and\ \citenamefont
  {Jarillo-Herrero}}]{Park2021b}%
  \BibitemOpen
  \bibfield  {author} {\bibinfo {author} {\bibfnamefont {J.~M.}\ \bibnamefont
  {Park}}, \bibinfo {author} {\bibfnamefont {Y.}~\bibnamefont {Cao}}, \bibinfo
  {author} {\bibfnamefont {L.}~\bibnamefont {Xia}}, \bibinfo {author}
  {\bibfnamefont {S.}~\bibnamefont {Sun}}, \bibinfo {author} {\bibfnamefont
  {K.}~\bibnamefont {Watanabe}}, \bibinfo {author} {\bibfnamefont
  {T.}~\bibnamefont {Taniguchi}}, \ and\ \bibinfo {author} {\bibfnamefont
  {P.}~\bibnamefont {Jarillo-Herrero}},\ }\href {\doibase
  10.48550/ARXIV.2112.10760} {\  (\bibinfo {year} {2021}{\natexlab{b}}),\
  10.48550/ARXIV.2112.10760}\BibitemShut {NoStop}%
\bibitem [{\citenamefont {Cao}\ \emph {et~al.}(2016)\citenamefont {Cao},
  \citenamefont {Luo}, \citenamefont {Fatemi}, \citenamefont {Fang},
  \citenamefont {Sanchez-Yamagishi}, \citenamefont {Watanabe}, \citenamefont
  {Taniguchi}, \citenamefont {Kaxiras},\ and\ \citenamefont
  {Jarillo-Herrero}}]{Cao2016}%
  \BibitemOpen
  \bibfield  {author} {\bibinfo {author} {\bibfnamefont {Y.}~\bibnamefont
  {Cao}}, \bibinfo {author} {\bibfnamefont {J.~Y.}\ \bibnamefont {Luo}},
  \bibinfo {author} {\bibfnamefont {V.}~\bibnamefont {Fatemi}}, \bibinfo
  {author} {\bibfnamefont {S.}~\bibnamefont {Fang}}, \bibinfo {author}
  {\bibfnamefont {J.~D.}\ \bibnamefont {Sanchez-Yamagishi}}, \bibinfo {author}
  {\bibfnamefont {K.}~\bibnamefont {Watanabe}}, \bibinfo {author}
  {\bibfnamefont {T.}~\bibnamefont {Taniguchi}}, \bibinfo {author}
  {\bibfnamefont {E.}~\bibnamefont {Kaxiras}}, \ and\ \bibinfo {author}
  {\bibfnamefont {P.}~\bibnamefont {Jarillo-Herrero}},\ }\href {\doibase
  10.1103/PhysRevLett.117.116804} {\bibfield  {journal} {\bibinfo  {journal}
  {Phys. Rev. Lett.}\ }\textbf {\bibinfo {volume} {117}},\ \bibinfo {pages}
  {116804} (\bibinfo {year} {2016})}\BibitemShut {NoStop}%
\bibitem [{\citenamefont {Uchida}\ \emph {et~al.}(2014)\citenamefont {Uchida},
  \citenamefont {Furuya}, \citenamefont {Iwata},\ and\ \citenamefont
  {Oshiyama}}]{Uchida2014}%
  \BibitemOpen
  \bibfield  {author} {\bibinfo {author} {\bibfnamefont {K.}~\bibnamefont
  {Uchida}}, \bibinfo {author} {\bibfnamefont {S.}~\bibnamefont {Furuya}},
  \bibinfo {author} {\bibfnamefont {J.-I.}\ \bibnamefont {Iwata}}, \ and\
  \bibinfo {author} {\bibfnamefont {A.}~\bibnamefont {Oshiyama}},\ }\href
  {\doibase 10.1103/PhysRevB.90.155451} {\bibfield  {journal} {\bibinfo
  {journal} {Phys. Rev. B}\ }\textbf {\bibinfo {volume} {90}},\ \bibinfo
  {pages} {155451} (\bibinfo {year} {2014})}\BibitemShut {NoStop}%
\bibitem [{\citenamefont {van Wijk}\ \emph {et~al.}(2015)\citenamefont {van
  Wijk}, \citenamefont {Schuring}, \citenamefont {Katsnelson},\ and\
  \citenamefont {Fasolino}}]{VanWijk2015}%
  \BibitemOpen
  \bibfield  {author} {\bibinfo {author} {\bibfnamefont {M.~M.}\ \bibnamefont
  {van Wijk}}, \bibinfo {author} {\bibfnamefont {A.}~\bibnamefont {Schuring}},
  \bibinfo {author} {\bibfnamefont {M.~I.}\ \bibnamefont {Katsnelson}}, \ and\
  \bibinfo {author} {\bibfnamefont {A.}~\bibnamefont {Fasolino}},\ }\href
  {\doibase 10.1088/2053-1583/2/3/034010} {\bibfield  {journal} {\bibinfo
  {journal} {2D Mater.}\ }\textbf {\bibinfo {volume} {2}},\ \bibinfo {pages}
  {034010} (\bibinfo {year} {2015})}\BibitemShut {NoStop}%
\bibitem [{\citenamefont {Koshino}\ \emph {et~al.}(2018)\citenamefont
  {Koshino}, \citenamefont {Yuan}, \citenamefont {Koretsune}, \citenamefont
  {Ochi}, \citenamefont {Kuroki},\ and\ \citenamefont {Fu}}]{Koshino2018}%
  \BibitemOpen
  \bibfield  {author} {\bibinfo {author} {\bibfnamefont {M.}~\bibnamefont
  {Koshino}}, \bibinfo {author} {\bibfnamefont {N.~F.~Q.}\ \bibnamefont
  {Yuan}}, \bibinfo {author} {\bibfnamefont {T.}~\bibnamefont {Koretsune}},
  \bibinfo {author} {\bibfnamefont {M.}~\bibnamefont {Ochi}}, \bibinfo {author}
  {\bibfnamefont {K.}~\bibnamefont {Kuroki}}, \ and\ \bibinfo {author}
  {\bibfnamefont {L.}~\bibnamefont {Fu}},\ }\href {\doibase
  10.1103/PhysRevX.8.031087} {\bibfield  {journal} {\bibinfo  {journal} {Phys.
  Rev. X}\ }\textbf {\bibinfo {volume} {8}},\ \bibinfo {pages} {031087}
  (\bibinfo {year} {2018})}\BibitemShut {NoStop}%
\bibitem [{\citenamefont {Chittari}\ \emph {et~al.}(2018)\citenamefont
  {Chittari}, \citenamefont {Leconte}, \citenamefont {Javvaji},\ and\
  \citenamefont {Jung}}]{Chittari2018}%
  \BibitemOpen
  \bibfield  {author} {\bibinfo {author} {\bibfnamefont {B.~L.}\ \bibnamefont
  {Chittari}}, \bibinfo {author} {\bibfnamefont {N.}~\bibnamefont {Leconte}},
  \bibinfo {author} {\bibfnamefont {S.}~\bibnamefont {Javvaji}}, \ and\
  \bibinfo {author} {\bibfnamefont {J.}~\bibnamefont {Jung}},\ }\href {\doibase
  10.1088/2516-1075/aaead3} {\bibfield  {journal} {\bibinfo  {journal}
  {Electron. Struct.}\ }\textbf {\bibinfo {volume} {1}},\ \bibinfo {pages}
  {015001} (\bibinfo {year} {2018})}\BibitemShut {NoStop}%
\bibitem [{\citenamefont {Leconte}\ \emph {et~al.}(2017)\citenamefont
  {Leconte}, \citenamefont {Jung}, \citenamefont {Leb\`egue},\ and\
  \citenamefont {Gould}}]{Leconte2017}%
  \BibitemOpen
  \bibfield  {author} {\bibinfo {author} {\bibfnamefont {N.}~\bibnamefont
  {Leconte}}, \bibinfo {author} {\bibfnamefont {J.}~\bibnamefont {Jung}},
  \bibinfo {author} {\bibfnamefont {S.}~\bibnamefont {Leb\`egue}}, \ and\
  \bibinfo {author} {\bibfnamefont {T.}~\bibnamefont {Gould}},\ }\href
  {\doibase 10.1103/PhysRevB.96.195431} {\bibfield  {journal} {\bibinfo
  {journal} {Phys. Rev. B}\ }\textbf {\bibinfo {volume} {96}},\ \bibinfo
  {pages} {195431} (\bibinfo {year} {2017})}\BibitemShut {NoStop}%
\bibitem [{\citenamefont {Yankowitz}\ \emph {et~al.}(2018)\citenamefont
  {Yankowitz}, \citenamefont {Jung}, \citenamefont {Laksono}, \citenamefont
  {Leconte}, \citenamefont {Chittari}, \citenamefont {Watanabe}, \citenamefont
  {Taniguchi}, \citenamefont {Adam}, \citenamefont {Graf},\ and\ \citenamefont
  {Dean}}]{Yankowitz2018}%
  \BibitemOpen
  \bibfield  {author} {\bibinfo {author} {\bibfnamefont {M.}~\bibnamefont
  {Yankowitz}}, \bibinfo {author} {\bibfnamefont {J.}~\bibnamefont {Jung}},
  \bibinfo {author} {\bibfnamefont {E.}~\bibnamefont {Laksono}}, \bibinfo
  {author} {\bibfnamefont {N.}~\bibnamefont {Leconte}}, \bibinfo {author}
  {\bibfnamefont {B.~L.}\ \bibnamefont {Chittari}}, \bibinfo {author}
  {\bibfnamefont {K.}~\bibnamefont {Watanabe}}, \bibinfo {author}
  {\bibfnamefont {T.}~\bibnamefont {Taniguchi}}, \bibinfo {author}
  {\bibfnamefont {S.}~\bibnamefont {Adam}}, \bibinfo {author} {\bibfnamefont
  {D.}~\bibnamefont {Graf}}, \ and\ \bibinfo {author} {\bibfnamefont {C.~R.}\
  \bibnamefont {Dean}},\ }\href {\doibase 10.1038/s41586-018-0107-1} {\bibfield
   {journal} {\bibinfo  {journal} {Nature}\ }\textbf {\bibinfo {volume}
  {557}},\ \bibinfo {pages} {404} (\bibinfo {year} {2018})}\BibitemShut
  {NoStop}%
\bibitem [{\citenamefont {Jung}\ and\ \citenamefont
  {MacDonald}(2013)}]{Jung2013}%
  \BibitemOpen
  \bibfield  {author} {\bibinfo {author} {\bibfnamefont {J.}~\bibnamefont
  {Jung}}\ and\ \bibinfo {author} {\bibfnamefont {A.~H.}\ \bibnamefont
  {MacDonald}},\ }\href {\doibase 10.1103/PhysRevB.88.075408} {\bibfield
  {journal} {\bibinfo  {journal} {Phys. Rev. B}\ }\textbf {\bibinfo {volume}
  {88}},\ \bibinfo {pages} {075408} (\bibinfo {year} {2013})}\BibitemShut
  {NoStop}%
\bibitem [{\citenamefont {Xiao}\ \emph {et~al.}(2010)\citenamefont {Xiao},
  \citenamefont {Chang},\ and\ \citenamefont {Niu}}]{Xiao2010b}%
  \BibitemOpen
  \bibfield  {author} {\bibinfo {author} {\bibfnamefont {D.}~\bibnamefont
  {Xiao}}, \bibinfo {author} {\bibfnamefont {M.-C.}\ \bibnamefont {Chang}}, \
  and\ \bibinfo {author} {\bibfnamefont {Q.}~\bibnamefont {Niu}},\ }\href
  {\doibase 10.1103/RevModPhys.82.1959} {\bibfield  {journal} {\bibinfo
  {journal} {Rev. Mod. Phys.}\ }\textbf {\bibinfo {volume} {82}},\ \bibinfo
  {pages} {1959} (\bibinfo {year} {2010})}\BibitemShut {NoStop}%
\end{thebibliography}%

\end{document}